\documentclass{article}

\usepackage{template_stuff/arxiv}

\usepackage[utf8]{inputenc} % allow utf-8 input
\usepackage[T1]{fontenc}    % use 8-bit T1 fonts
\usepackage{hyperref}       % hyperlinks
\usepackage{url}            % simple URL typesetting
\usepackage{booktabs}       % professional-quality tables
\usepackage{amsfonts}       % blackboard math symbols
\usepackage{nicefrac}       % compact symbols for 1/2, etc.
\usepackage{microtype}      % microtypography
\usepackage{graphicx}
\usepackage[style=numeric-comp,backend=biber,giveninits=true,sorting=none,maxnames=99,minnames=99, url=false,date=year]{biblatex}
\AtEveryBibitem{
    \clearfield{issn}
    \clearfield{language}
    \clearlist{language}
    \clearfield{primaryClass}
    \clearlist{primaryClass}
    \clearfield{archivePrefix}
    \clearlist{archivePrefix}
    \clearfield{eprint}
    \clearlist{eprint}
}
\usepackage{doi}
\usepackage{placeins} % for floatbarrier etc

\usepackage{siunitx}
\usepackage{amsmath}
\usepackage{amssymb}
\usepackage{xspace}
\usepackage{authblk}
\usepackage{makecell}

\usepackage{bm}
\usepackage{subfig}  % for subfigures
\usepackage{caption}  % for subfigures
\usepackage[framemethod=tikz]{mdframed}
\usetikzlibrary{calc}
\usetikzlibrary{positioning} 
\usetikzlibrary {math} 
\usetikzlibrary {arrows.meta} 
\usepackage{enumitem} % for costumizing items
\usepackage{xcolor}
\usepackage{soul}
\usepackage[normalem]{ulem}
\newcommand{\comment}[1]{}
\newmdenv[
    topline=false,
    bottomline=false,
    rightline=false,
    innerrightmargin=0pt
]{siderule}
    {\begin{siderule}\textbf{Note:}}
    {\end{siderule}}
\newcommand{\eeq}{\ .}  % end of equation
\newcommand{\ceq}{\ ,}  % comma in equation
\newcommand{\bi}{\begin{itemize}[topsep=-5pt, itemsep=1pt, parsep=0pt, partopsep=0pt]}
\newcommand{\ei}{\end{itemize}\vspace{5pt}}

\newcommand{\mylog}{\log \!}

\newcommand{\Exp}[1]{{\mathbb{E}}\!\left[#1\right]}

\newcommand{\Var}[1]{{\text{Var}}\!\left(#1\right)}
\newcommand{\Ent}[1]{{\text{H}}\!\left(#1\right)}
\newcommand{\Dkl}[2]{{D}_{\text{KL}}\!\left(#1 \,||\, #2\right)}
\newcommand{\Inf}[1]{{\text{I}}\!\left(#1\right)}
\newcommand{\IG}[1]{{\text{IG}}\!\left(#1\right)}

\newcommand{\RIIG}[1]{{\text{RIIG}}\!\left(#1\right)}
\newcommand{\M}{\mathcal{M}}

\newcommand{\N}{\mathcal{N}}
\newcommand{\obs}{\text{obs}}
\newcommand{\true}{\text{true}}
\newcommand{\fields}{\text{fields}}
\newcommand{\SNR}{\text{SNR}}
\newcommand{\noise}{\boldsymbol{\varepsilon}}
\newcommand{\bmu}{\boldsymbol{\mu}}
\newcommand{\ba}{\boldsymbol{a}}
\newcommand{\bb}{\boldsymbol{b}}

\newcommand{\f}{\boldsymbol{f}}
\newcommand{\A}{\boldsymbol{A}}
\newcommand{\partialfy}[2]{\dfrac{\partial \f_#1}{\partial \by_#2}}
\newcommand{\vphantompartial}{\vphantom{\partialfy{1}{1}}}

\newcommand{\bd}{\boldsymbol{d}}
\newcommand{\bp}{\boldsymbol{p}}

\newcommand{\bx}{\boldsymbol{x}}
\newcommand{\by}{\boldsymbol{y}}
\newcommand{\bc}{\boldsymbol{c}}

\newcommand{\dd}{\,\mathrm{d}} % nice integral d
\newcommand{\diag}{\text{diag}\!}

\newcommand{\greytext}[1]{\textcolor[rgb]{0.2,0.2,0.2}{#1}}
\newcommand{\greybf}[1]{\greytext{\textbf{#1}}}

\definecolor{mygreen}{HTML}{56A200}
\setstcolor{red}
\newcommand{\delete}[1]{}
\newcommand{\new}[1]{#1}
\title{Multi-Physics-Enhanced Bayesian Inverse Analysis:\\Information Gain from Additional Fields}

\author[1,*]{Lea J.~Haeusel}
\author[1]{Jonas Nitzler}
\author[1]{Lea J.~Köglmeier}
\author[1]{Wolfgang A.~Wall}

\affil[1]{
    Institute for Computational Mechanics\\
    Technical University of Munich\\
    Boltzmannstr. 15\\
    85748 Garching b. München\\
    Germany
}
\affil[ ]{}
\affil[*]{
    Corresponding author: \texttt{lea.haeusel@tum.de}
}

\renewcommand{\shorttitle}{Multi-Physics-Enhanced Bayesian Inverse Analysis}

\hypersetup{
pdftitle={Multi-Physics-Enhanced Bayesian Inverse Analysis: Information Gain from Additional Fields},
pdfsubject={multi-physics-enhanced Bayesian inverse analysis, information gain, multi-physics data, multi-physics computational modeling, coupled fields},
pdfauthor={Lea J.~Haeusel, Jonas Nitzler, Lea J.~Köglmeier, Wolfgang A.~Wall},
pdfkeywords={Multi-physics-enhanced Bayesian inverse analysis, Multi-physics models, Multi-physics data, Coupled fields, Information gain},
}

\begin{document}
\maketitle

%%%%%%%%%%%%%%%%%%%%%%%%%%%%%%%%%%%%%%%%%%%%%%%%%%%%%%%%%%%%%%%%%%%%%%%%%%%%%%%%%%%%%%%%%%%%%%%%%%%
\begin{abstract}
    \new{
    % The issue
    Inverse analysis, such as model calibration, often suffers from a lack of informative data in complex real-world scenarios.
    % The gap
    The standard remedy, designing new experimental setups, is often costly and time-consuming, while readily available but seemingly useless data are ignored.}

    % Our idea
    \new{This work proposes incorporating such data from additional physical fields into the inverse analysis, even when the forward model solves a single-physics problem.
    A Bayesian framework easily incorporates the additional data and quantifies the resulting uncertainty reduction.}

    % What have we done
    \new{We formally introduce the proposed method, which we denote as multi-physics-enhanced Bayesian inverse analysis.
    Moreover, this work is the first to quantify the reduction in parameter uncertainty by comparing the information gain from the prior to the posterior when using single-physics versus multi-physics data.}

    % Our results
    \new{We demonstrate the potential of the proposed method in two exemplary applications.}
    Our results show that even a few or noisy data points from an additional physical field can considerably increase the information gain, even \new{when} the physical field is only weakly or one-way coupled.
    
    % Conclusion / outlook
    \new{Overall, this} work proposes and promotes the future use of multi-physics-enhanced Bayesian inverse analysis as a cost- and time-saving game-changer across various fields of science and industry, particularly in medicine.
\end{abstract}
%%%%%%%%%%%%%%%%%%%%%%%%%%%%%%%%%%%%%%%%%%%%%%%%%%%%%%%%%%%%%%%%%%%%%%%%%%%%%%%%%%%%%%%%%%%%%%%%%%%

\keywords{Bayesian inverse analysis \and Model calibration \and Multi-physics data \and Coupled fields \and Information gain}

%%%%%%%%%%%%%%%%%%%%%%%%%%%%%%%%%%%%%%%%%%%%%%%%%%%%%%%%%%%%%%%%%%%%%%%%%%%%%%%%%%%%%%%%%%%%%%%%%%%
%%%%%%%%%%%%%%%%%%%%%%%%%%%%%%%%%%%%%%%%%%%%%%%%%%%%%%%%%%%%%%%%%%%%%%%%%%%%%%%%%%%%%%%%%%%%%%%%%%%
\section{Introduction}

Model calibration, which is the identification of model parameters, remains one of the most challenging tasks in all fields of the applied sciences, especially in medicine, and in various industrial applications.
The challenges lie in formulating the problem properly, in the need for accurate, efficient, and robust methods, as well as in the availability and quality of the data. 
The standard approach to identify parameters from experiments and gain quantitative insight is to conduct a specific experiment and use the resulting measurements of obviously relevant quantities.
Many excellent examples and powerful methods for this approach can be found in the literature, particularly in the field of computational model calibration, the field on which our work focuses.
However, as \delete{research}\new{inverse} problems \delete{and computational models }have become increasingly complex, this standard approach often faces strict limitations on the type of experiments and measurements that can be conducted at a reasonable cost.
This \delete{development }has \delete{led to}\new{added to the already substantial}\delete{great  difficulties} \new{challenges} in estimating model parameters.
The standard response has been to design new experiments, which are often very expensive and time-consuming.
In contrast, a vast number of measurements are \new{often }readily available, or could be obtained easily and cheaply\delete{, but}\new{. These measurements} are typically neglected because they are not directly linked to the quantity of interest.
In this work, we propose an approach that avoids additional costly and time-consuming experiments and instead leverages \new{this }data from seemingly unrelated fields.

We illustrate our approach with a simple example.
Suppose the task is to take measurements in a certain section of a river to infer some unknown flow boundary conditions.
Only very few flow sensors can be placed in the river, which makes it extremely difficult to identify the boundary conditions.
Instead of buying additional flow sensors and placing them in hard-to-reach locations, one could \delete{simply pour some color into the river.}\new{observe the river's transport of a substance that}\delete{
While the color} does not affect the flow field\new{, e.g., the transport of color.}\delete{, even} \new{Even} a few inexpensive, low-resolution images of the \new{substance or} color distribution can greatly improve the identification of the boundary conditions.
\delete{An equivalent approach is standard in gas flows, where smoke is added to a colorless gas to learn more about the flow field.
}Our approach in this work is similar to this simple idea.
We propose not to waste but to use data from additional fields, even if \new{these fields are irrelevant for the forward simulation task and even if} they initially seem irrelevant for the underlying inverse problem.
In our simple example above, the (forward) solution of the river flow field from the boundary conditions only asks for solving the Navier-Stokes equations for the flow velocity, which is a single field.
However, adding a transported scalar field \new{for the inverse analysis} allows us to also learn \new{these boundary conditions }from the \new{substance or }color distribution\delete{, which is highly valuable in the inverse analysis}.

\delete{Another}\new{A particular} challenge for the inverse problem of model calibration is the ill-posedness \delete{when it is conducted deterministically}\new{in deterministic formulations}.
This ill-posedness arises due to \new{several reasons. On the one hand,}\delete{the fact that} every model is merely an approximation of the real system, which the observations capture in a noisy manner.
\new{As a result, an exact solution generally does not exist for a chosen approximate model and the available measurements.
On the other hand, many forward problems governed by partial differential equations have an integral or smoothing character.
The corresponding inverse operator then behaves like a differentiation, which strongly amplifies noise in the observations.
As a consequence, even small perturbations in the observations can lead to large changes in the inferred parameters.
Deterministic inverse methods address this ill-posedness by minimizing a discrepancy measure between model outputs and observations, while adding some form of regularization.
Although regularization can help, deterministic inverse problems typically still face the existence of multiple minima and unstable solutions with respect to small changes in the observations.}\delete{This causes the existence of multiple minima, and even small perturbations in the observations can lead to large changes in the inferred parameters.}
\new{The inverse problem may therefore remain ill-posed with deterministic methods.}
Moreover, \delete{the insight provided by deterministic approaches is limited}\new{deterministic approaches provide limited insight} as they yield only a single value of the parameters \new{and do not quantify the uncertainty of this estimate}.
A probabilistic alternative to deterministic inverse analysis is Bayesian inverse analysis.
In contrast to the deterministic inverse problem, the Bayesian inverse problem is \delete{inherently }well-posed in the probabilistic sense \new{for most practical applications}.
\delete{It is also a lot more informative as it provides a full probability distribution over all possible parameters, which quantifies the remaining uncertainty.}\new{By representing the unknown parameters as a random variable, the Bayesian approach yields a probability distribution over all possible parameter values.
This probability distribution is a lot more informative than a deterministically derived point estimate and quantifies the remaining uncertainty in the parameters.}
Most importantly for this work, Bayesian inverse analysis also easily integrates observations from multiple sources or physical fields and can account for different noise levels across these fields.

\begin{figure}[]
    \hspace{-0.45cm}
    \begin{tikzpicture}
        \node at (0.0,0) [anchor=center] {
            \begin{tabular}{c}
            \quad \ \small\textbf{Noisy observations} \\[3pt]
            \includegraphics[height=0.18\textwidth]{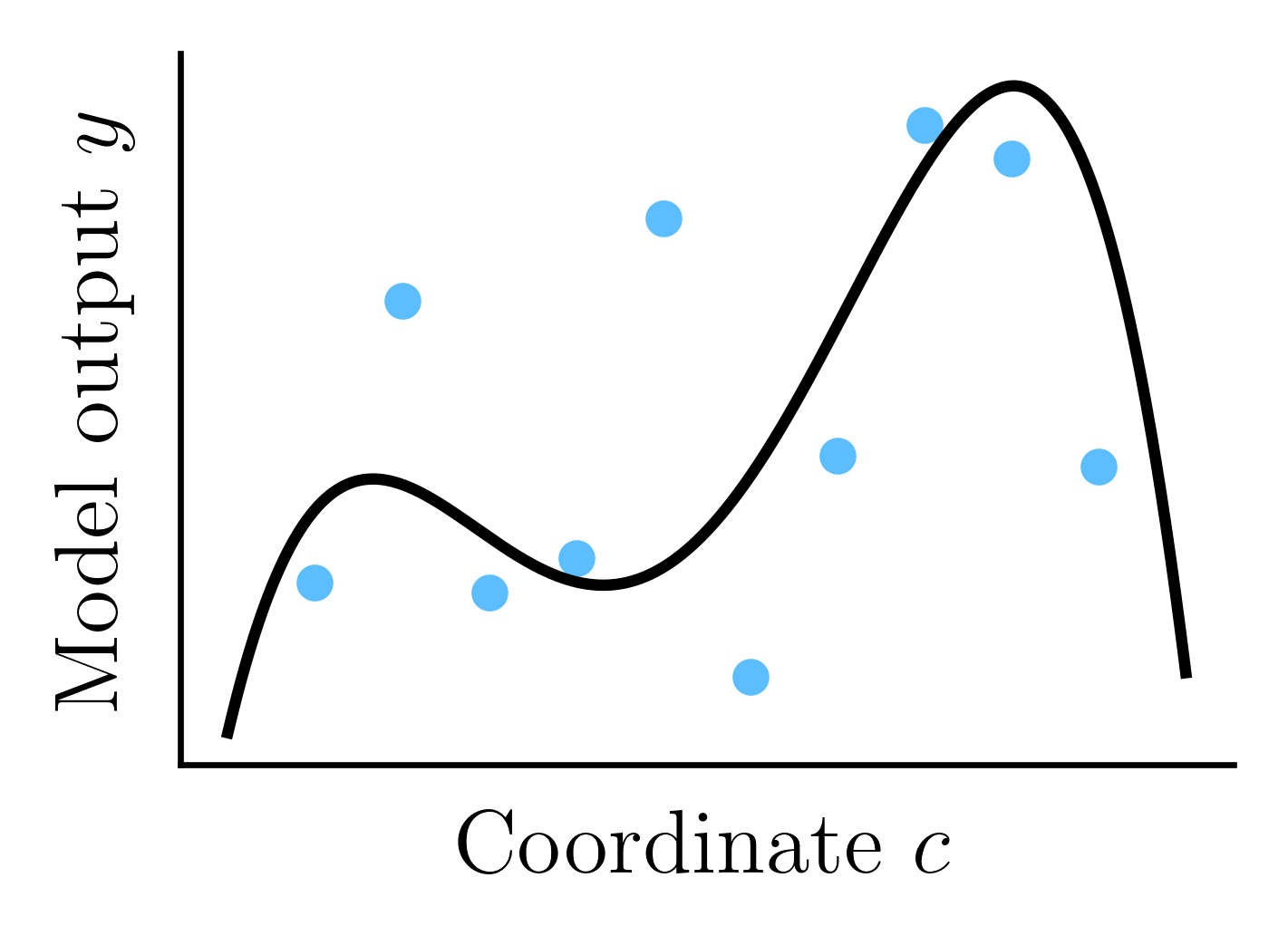}
            \end{tabular}
        };
        \node at (4.2,0) [anchor=center] {
            \begin{tabular}{c}
            \quad \ \small\textbf{Scarce observations} \\[3pt]
            \includegraphics[height=0.18\textwidth]{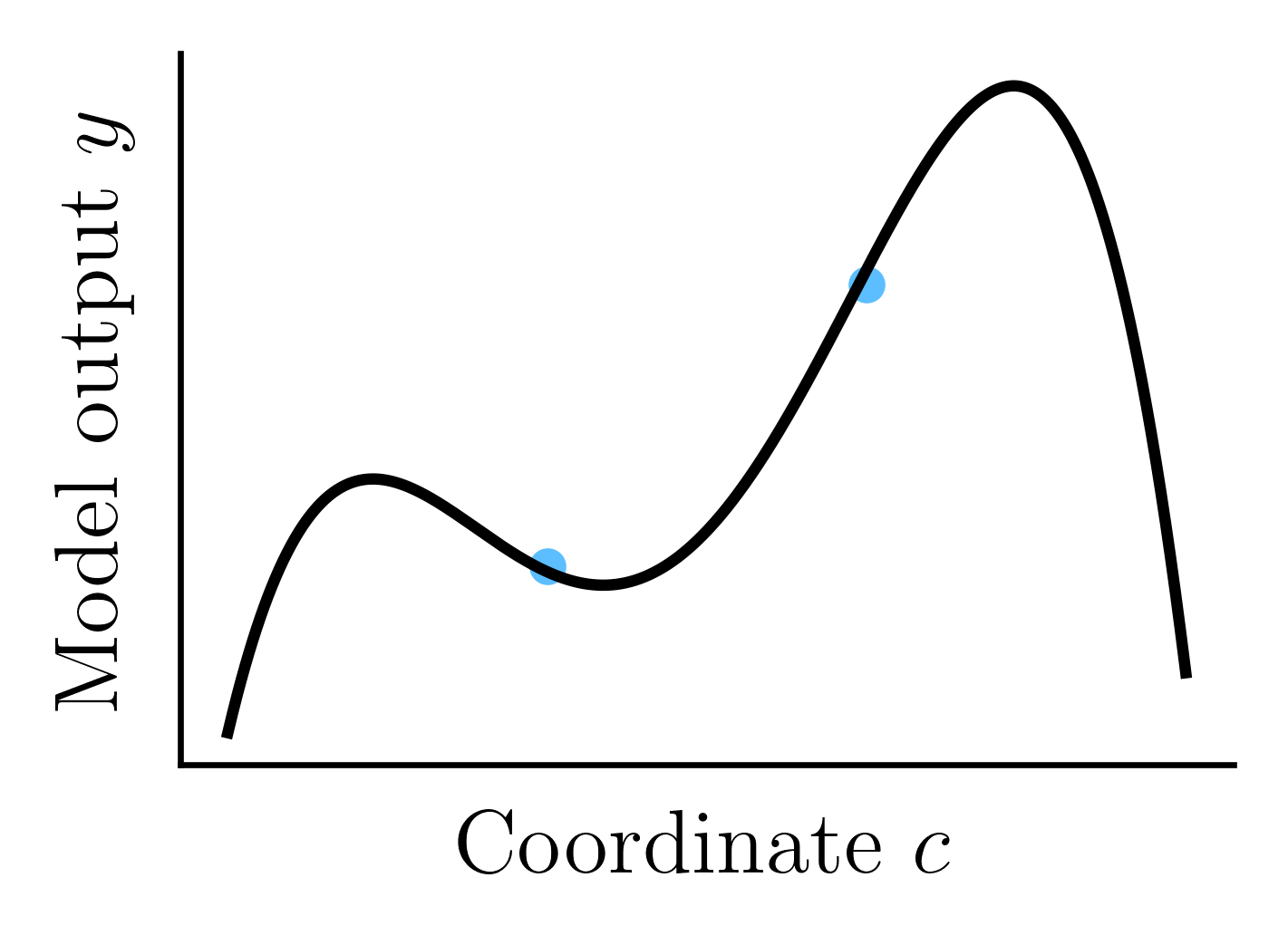}
            \end{tabular}
        };
        \node at (9.0,0) [anchor=center] {
            \begin{tabular}{c}
            \quad \ \small\textbf{Integral observations} \hphantom{mmmm} \\[3pt]
            \includegraphics[height=0.18\textwidth]{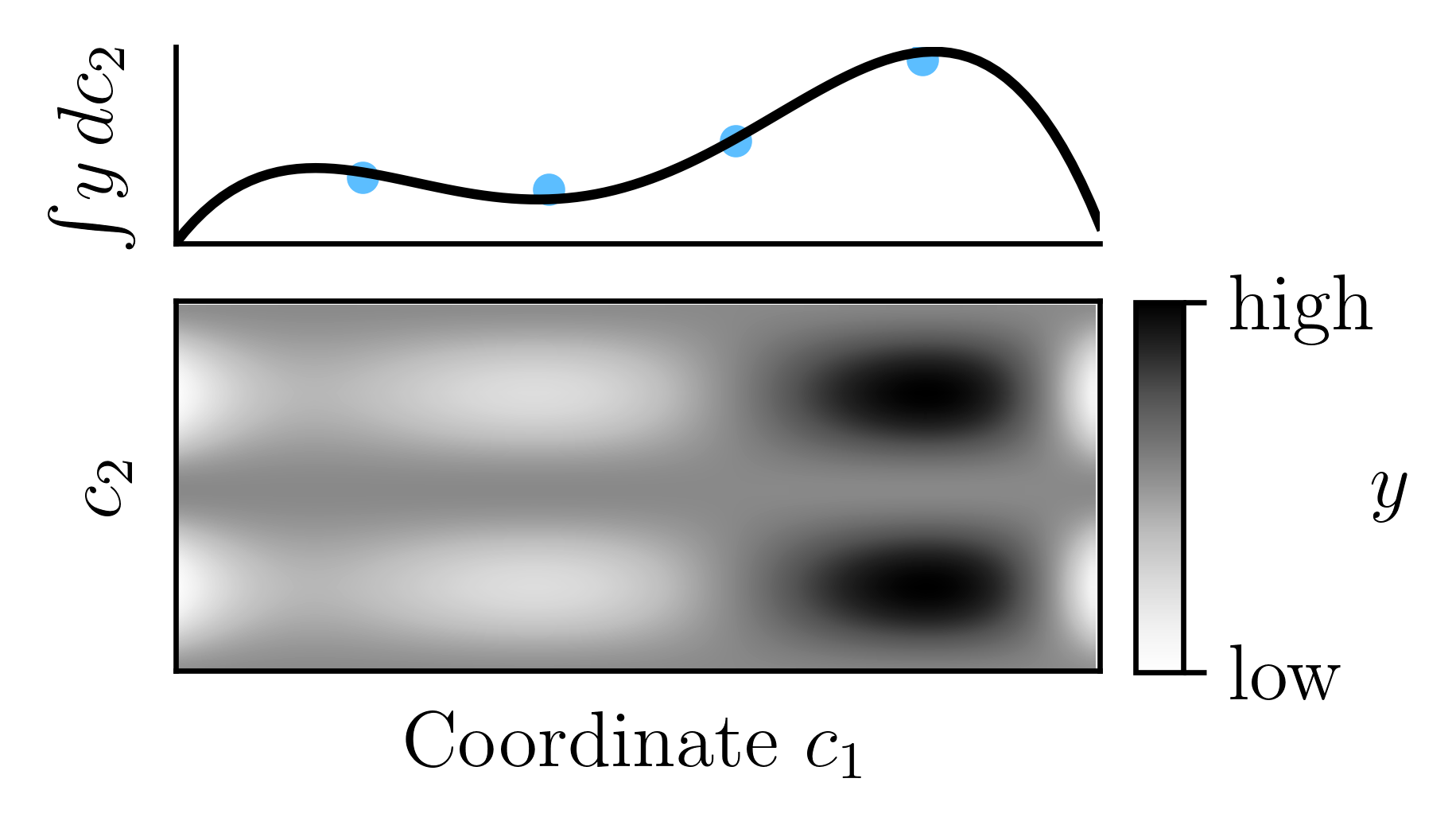}
            \end{tabular}
        };
        \node at (13.1,0.9) [anchor=center] {
            \includegraphics[width=0.18\textwidth]{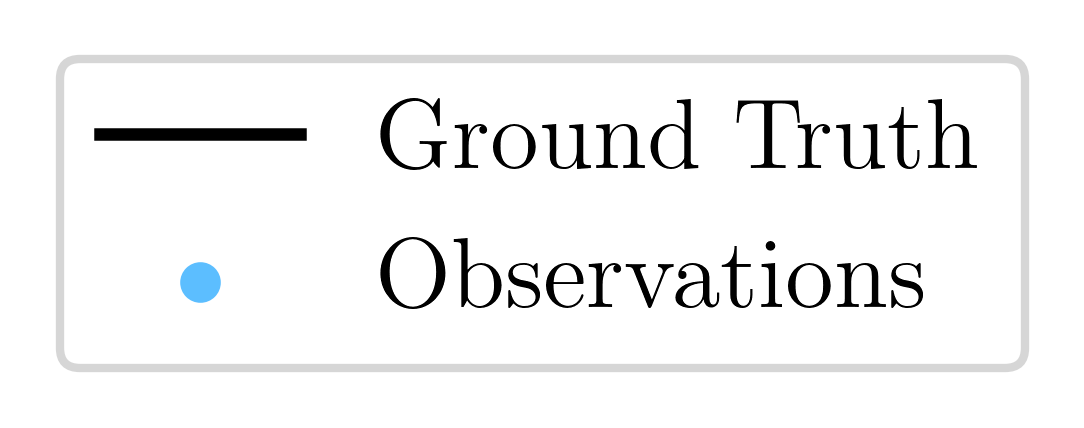}
        };
    \end{tikzpicture}
    \vspace{-15pt}
    \caption{Three types of limitations in the observations of a scalar model output $y = \M(\bx)$. The resulting observations may fail to sufficiently constrain the parameters $\bx$ when performing Bayesian inverse analysis.}
    \label{fig:observational_limitations}
\end{figure}

In this paper, we propose a Bayesian inverse analysis approach that adds an auxiliary, coupled field for the identification of computational model parameters.
By computational models, we refer to discretized, physics-based systems described by (partial) differential equations.
An evaluation of such a computational model solves these governing equations numerically, for example, using the finite element method.
From a higher-level perspective, we view the computational model as a deterministic function $\M$ that maps the (model) parameters $\bx$ to model outputs $\by$:
\begin{align}
    \M(\bx) = \by \eeq
    \label{eq:model}
\end{align}
Note that we use lower-case bold symbols to indicate vectors and lower-case regular symbols to indicate scalars.
As model parameters \new{$\bx$} are often hard to measure directly, the inverse analysis of model calibration derives them from measurements of $\by$.
% Introduce inverse analysis
In such cases, Bayesian inverse analysis offers a powerful framework to both include and update a prior belief $p(\bx)$ about the parameters $\bx$ based on \new{noisy} observations $\by_\obs$ of the model outputs $\by$.
The result of this inverse analysis is the posterior probability distribution $p(\bx | \by_\obs)$ according to Bayes' theorem:
\begin{align}
    \underbrace{p(\bx | \by_\obs)}_{\text{posterior}}
     & = \frac{
        \overbrace{p(\by_\obs|\bx)}^{\text{likelihood}}
        \overbrace{p(\bx)}^{\text{prior}}
    }
    {\int p(\by_\obs|\bx) p(\bx) d\bx} \eeq
    \label{eq:bayes_rule}
  \end{align}
The posterior distribution can remain uninformative, meaning it stays similar to the prior distribution with little change in uncertainty.
An uninformative posterior distribution can occur even if we choose an appropriate prior distribution, likelihood model, inverse method, and computational model \cite{hervas2023}.
In this case, the observations $\by_\obs$ do not sufficiently constrain the range of probable parameters $\bx$.
On the one hand, this can result from observations that are noisy or sparse, as depicted in Figure \ref{fig:observational_limitations}.
Limited data availability and quality are common challenges across all fields of science and engineering applications.
For example, it impedes patient-specific predictions in medical applications.
On the other hand, insufficient constraints, and thus an infinite number of solutions for the parameters $\bx$, can also appear irrespective of the number and noise of the observations.
\delete{For example}\new{To provide an academic example}, the simple model $y = \M(\bx=[x_1, x_2]^T) = t\cdot(x_1 + x_2)$ will never result in a unique solution for \new{the correlated parameters} $x_1$ and $x_2$ when observing $y$ over time $t$.
This challenge also arises \new{in practical applications} when \delete{only integral quantities are observable}\new{the model output $\by$ can only be observed as its integral over a domain}, as illustrated in Figure \ref{fig:observational_limitations}.
For instance, in respiratory medicine, only the integral airflow at the trachea can be easily measured.
Local airflow in smaller airways remains inaccessible with current non-invasive clinical methods.

%%%%%%%%%%%%
%%%%%%%%%%%%
  
%%%%%%%%%%%%
\begin{figure}[b]
    \centering
    \vspace{-5pt}
    \includegraphics[scale=0.07]{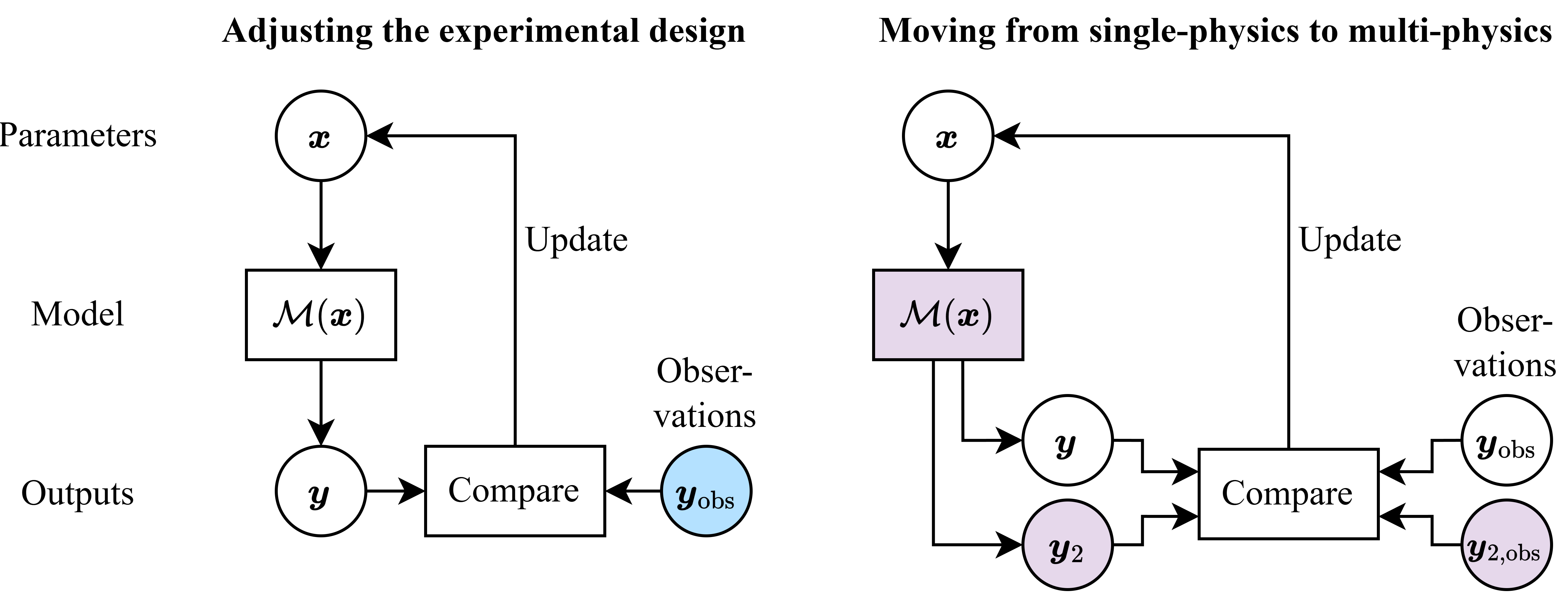}
    \caption{Two approaches to enhancing an inverse analysis via the observations.
    Components changed or added by each approach are highlighted in color.
    \emph{Left:} The first approach, adjusting the experimental design, changes the first-field observations $\by_\obs$.
    \emph{Right:} The second approach, moving from single-physics to multi-physics observations, leaves the first-field observations $\by_\obs$ unchanged and adds observations $\by_{2, \obs}$ of a second physical field.
    As a result, the model $\M$ needs to model both the original physical field $\by$ and the additional physical field $\by_2$.}
    \label{fig:inverse_improvements_comparison}
\end{figure}
%%%%%%%%%%%%

% Introduce current ways to improve inverse analysis
To address these observational challenges, we identify two approaches that reduce uncertainty in an uninformative posterior distribution by enhancing the observations.
% Introduce optimal experimental design
The first and more common approach is to optimize the experimental design, e.g., by improving the locations of the observations $\by_\obs$.
% Limitations optimal experimental design
While this approach is the current standard, adjusting real-life experiments is often laborious and costly.
In the context of Bayesian optimal experimental design, the optimization further comes with a high computational cost due to the integration over possible designs and repeated evaluation of the posterior distribution \cite{ryan2016review}.
What is more, optimizing the experimental design while observing the same physical field cannot overcome intrinsic observational limitations of this field (compare Figure \ref{fig:observational_limitations}).
% Introduce multi-physics approach
To overcome these limitations, observations of an additional physical field can provide valuable information and restrict the space of probable input parameters $\bx$.
% Define what we mean by field
By physical field or simply field, we refer to a physical quantity governed by a distinct set of physical laws.
Figure \ref{fig:inverse_improvements_comparison} illustrates the difference between this multi-physics-enhanced inverse approach and adjusting the experimental design for a generic inverse analysis.
To incorporate the multi-physics observations, the model $\M$ must be extended to model both the original physical field $\by$ and the additional physical field $\by_{2}$.
It is important to emphasize that this model extension does not aim to improve the forward model predictions.
Instead, it seeks to enhance the inverse analysis using the insights from comparing the new model outputs $\by_{2}$ with the additional observations $\by_{2, \obs}$.
Note that more than one physical field can be added in general.
 
\delete{While}\new{Upon presentation of this approach,} it may seem trivial that \new{incorporating }additional observations from coupled physical fields improve\new{s} a Bayesian inverse analysis\new{.}\delete{, this approach is still rarely applied in practice.
During the course of our work, we became aware of one noteworthy exception.
This exception is the application of multi-physics-enhanced Bayesian inverse analysis}
\new{Nevertheless, to our knowledge, this multi-physics-enhanced approach, including a quantification of the information gain, has not yet been directly proposed for the calibration of computational models.
We found only the application of a related idea in a different context, namely the approach of Bayesian joint inversion}
in the domain of geophysics \delete{under the term Bayesian joint inversion }\cite{rabben2008, jardani2010, bodin2012, afonso2013, shen2013, rosas2014, bodin2016, guo2016, agostinetti2018, blatter2019, barnoud2021, manassero2021, peng2021, yang2021, yao2023}.
However, in contrast to the \delete{context of this}\new{present} work, these Bayesian \new{joint inversions}\delete{inverse analyses} are not based on coupled but uncoupled fields and do not use models based on discretized differential equations.
To the best of our knowledge, only two studies have utilized the potential of multi-physics observations to enhance inverse analysis using a coupled computational model.
The first study added displacement observations to electrocardiographic observations to improve the inverse analysis of the electrical state of a heart \cite{corrado2015}.
This improvement is particularly noteworthy as the second, mechanical field was only weakly one-way coupled to the first, electrical field.
The second study\new{, coming from our own group,} probabilistically inferred the local stiffness variation in a ventilated human lung \cite{dinkel2024}.
\delete{The authors solved this}\new{In this study, the} inverse problem \new{was solved} by combining observations of the integral airflow at the ventilator, the first field, with voltage measurements from electrical impedance tomography (EIT), the second, one-way coupled field.
Without the additional EIT data containing spatially resolved information, the posterior distribution in this study would have remained uninformative.

This brief literature review demonstrates that the use of multi-physics observations to enhance Bayesian inverse analysis of computational models is far from being widely established.
We attribute this partially to the implementation costs of multi-physics models, but primarily to a lack of awareness of the potential of this approach.
Therefore, the goal of this work is to formally introduce the multi-physics-enhanced approach, to demonstrate its potential, and to encourage researchers and practitioners across various disciplines to adopt it.
We particularly propose to leverage the potential of multi-physics-enhanced Bayesian inverse analysis in any of the following scenarios\new{, starting from a single-physics inverse analysis}:
\begin{enumerate}
    \item Multi-physics observations are readily available, and a corresponding multi-physics model can be implemented with little effort, e.g., due to one-way coupled or uncoupled physical fields.
    \item A multi-physics model is readily available, and corresponding multi-physics observations are obtainable with little effort.
    \item First-field observations are not available or only with great difficulty, and second-field observations and a corresponding multi-physics model are available at a reasonable cost.
\end{enumerate}
In the first two scenarios, the small effort of adopting the multi-physics-enhanced approach can be rewarded with a substantial reduction in posterior uncertainty.
While the effort may be higher in the third scenario, the potential reward is even greater if the lack of first-field observations has so far prohibited the inverse analysis.
\new{Our contribution is to propose the multi-physics-enhanced approach for a broad set of fields and computational models.
}\delete{Besides proposing the multi-physics-enhanced approach, our contribution is to raise}\new{We further contribute by raising} awareness of \delete{its}\new{the} low cost-benefit ratio \new{of this approach} and \delete{to investigate}\new{by investigating} the uncertainty reduction through observations of an additional physical field.
Our work is the first to quantify this uncertainty reduction by computing the information gain from the prior to the posterior distribution.
In particular, we consider the relative increase in information gain when moving from single-physics to multi-physics-enhanced Bayesian inverse analysis.
We analyze this relative increase in information gain for different numbers of observations, noise levels, coupling types, and coupling strengths.

In Section \ref{sec:bia}, we briefly present our multi-physics-enhanced Bayesian inverse analysis framework.
The different coupling types and the notation of multi-physics models are defined in Section \ref{sec:coupled_multi-physics_models}.
Section \ref{sec:info_gain} introduces the relative increase in information gain as our measure of uncertainty reduction when adding observations of a second physical field.
We then analyze this relative increase in information gain for varying numbers of observations and signal-to-noise ratios in the second field using a scalar-valued, one-way coupled electromechanical model in Section \ref{sec:simple_model}.
In Section \ref{sec:poro_model}, we consider a finite-element-based, fully coupled porous medium model to compute the relative increase in information gain for varying signal-to-noise ratios in the first and second field and for varying coupling strengths.
Finally, we summarize our findings and draw conclusions regarding the potential of multi-physics-enhanced Bayesian inverse analysis in Section \ref{sec:conclusion}.

\FloatBarrier
%%%%%%%%%%%%%%%%%%%%%%%%%%%%%%%%%%%%%%%%%%%%%%%%%
\section{Bayesian inverse analysis with multi-physics observations}
\label{sec:bia}

We start with an introduction to Bayesian inverse analysis to perform parameter calibration based on computational models.
To include multi-physics observations in such \new{an analysis}\delete{analyses}, we propose a multi-physics log-likelihood function.
In Section \ref{sec:coupled_multi-physics_models}, we provide an overview of coupling types in multi-physics models and their implications for multi-physics-enhanced Bayesian inverse analysis.
To avoid unnecessary complexity, our notation does not distinguish between random variables, their realizations, and normal variables.

% Introduce probabilistic model calibration
Model calibration aims to identify model parameters $\bx$ that minimize the discrepancy between the model outputs $\by=\M(\bx)$ and observational data $\by_\obs$.
In the context of computational models, the model parameters $\bx$ can represent material parameters, geometric parameters, boundary conditions, and initial conditions.
Identifying the parameters $\bx$ from the observations $\by_\obs$ constitutes an inverse problem.
From a deterministic perspective, this inverse problem is formulated as an optimization problem, which is often ill-posed.
This ill-posedness arises due to the existence of multiple minima or because small perturbations in the observations $\by_\obs$ lead to large changes in the inferred parameters $\bx$.
Both effects may result from noise in the observations or from model error, since any model is just an approximation of the real system.
Moreover, deterministic solution approaches yield only a single value of the parameters $\bx$.
This point estimate may not be a global optimum and neglects alternative solutions without providing an uncertainty estimate.
A probabilistic alternative to deterministic inverse analysis is Bayesian inverse analysis.
Bayesian inverse analysis reformulates the inverse problem as a statistical inference problem according to Bayes' theorem (see Equation \ref{eq:bayes_rule}).
In contrast to the deterministic inverse problem, the Bayesian inverse problem is \delete{inherently }well-posed in a probabilistic sense \new{for most practical applications} and naturally accounts for observational noise and model error.
Its solution is a full probability distribution over all possible parameters $\bx$, which quantifies the remaining uncertainty in the parameters.
Most importantly for this work, Bayesian inverse analysis also easily integrates observations  $\by_\obs$ from multiple sources or physical fields.
We will demonstrate this in the following, after introducing the Bayesian approach in more detail.

\begin{figure}[]
    \centering
    \includegraphics[width=0.8\textwidth]{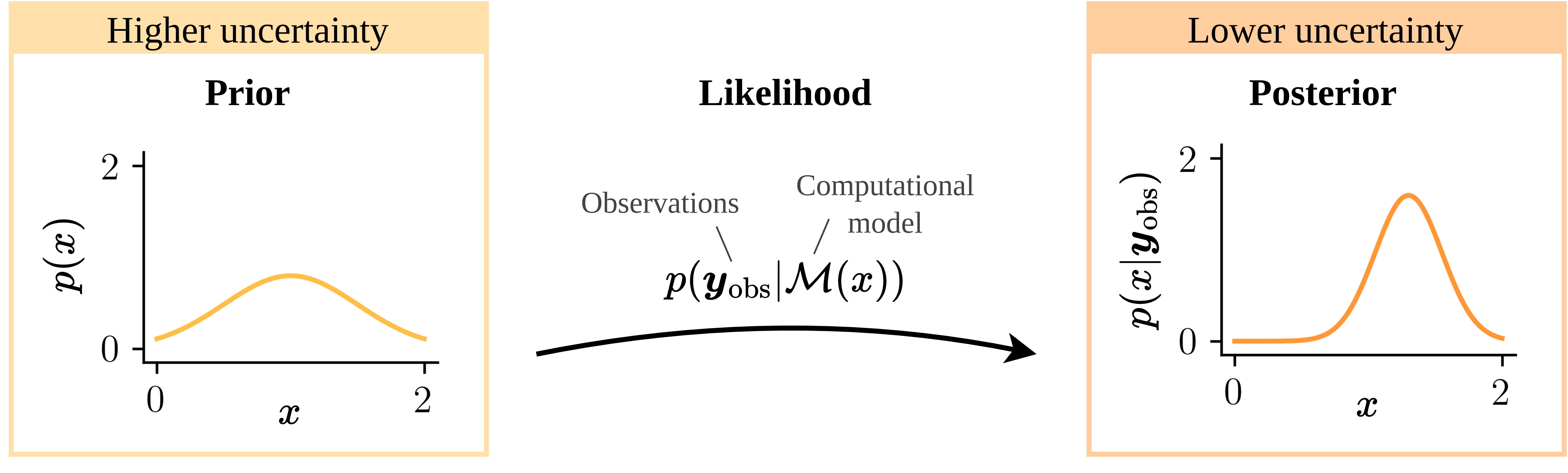}
    \caption{During Bayesian inverse analysis, the prior probability distribution $p(x)$ with higher uncertainty in the continuous scalar parameter of interest $x$ (\emph{left}) is updated using the likelihood $p(\by_\obs|\M(x))$. The likelihood expresses the probability of any observations $\by_\obs$ given an evaluation of the computational model $\M(x)$. The result of the analysis is the posterior probability distribution $p(x|\by_\obs)$ with lower uncertainty in the parameter $x$ (\emph{right}).}
    \label{fig:prior_to_posterior}
\end{figure}

% Introduction prob. distr., prior, posterior
In a Bayesian framework, the uncertain parameters $\bx$ are treated as random variables, and their uncertainty is represented by a probability distribution.
This probability distribution specifies the probability (for discrete parameters) or probability density (for continuous parameters) associated with each possible value of the parameters $\bx$.
As illustrated in Figure \ref{fig:prior_to_posterior}, the prior probability distribution $p(\bx)$ encodes our prior knowledge about the parameters $\bx$ before observing any data.
The Bayesian approach updates this prior belief by incorporating the observations $\by_\obs$.
This update is based on the likelihood $p(\by_\obs|\bx)$, which describes the probability distribution of observing any $\by_\obs$ given specific parameters $\bx$.
To assess this probability distribution, we need to map from the space of the parameters $\bx$ to the space of the observations $\by_\obs$.
The deterministic computational model $\by = \M(\bx)$ provides this mapping, modifying the likelihood to $p(\by_\obs|\bx) = p(\by_\obs|\M(\bx))$
\footnote{Please note that $p(\by_\obs|\bx) = p(\by_\obs|\M(\bx))$ only holds for deterministic models $\by = \M(\bx)$:\vspace{3pt}\\ $p(\by_\obs|\bx) = \int p(\by_\obs|\by) p(\by|\bx) d\by = \int p(\by_\obs|\by) \delta(\by - \M(\bx)) d\by = p(\by_\obs|\M(\bx))$.}.
Multiplying the prior and this likelihood, the posterior distribution $p(\bx|\by_\obs)$ results from Bayes' theorem given in Equation \ref{eq:bayes_rule}.
This posterior distribution $p(\bx|\by_\obs)$ finally reflects the remaining uncertainty in the parameters $\bx$ after incorporating the observations $\by_\obs$.

%%%%%%%%%%%%%%%%%%%%%%%%%%
%%%%%%%%%%%%%%%%%%%%%%%%%%

% Explain likelihood
During Bayesian inverse analysis, the likelihood distribution $p(\by_\obs|\M(\bx))$ is typically evaluated point-wise for some fixed observations $\by_\obs$.
In this case, the likelihood can be seen as a function of the parameters $\bx$.
An evaluation of this likelihood function for given parameters $\bx$ involves computing the model output $\by=\M(\bx)$ for this parameter and comparing it to the observations $\by_\obs$.
For observations of multiple physical fields, each of the $N_{\fields}$ physical fields is observed at $N_{\obs,j}$ spatio-temporal coordinates $\bc_{ij}$.
Here, $i$ and $j$ denote the $i$-th observation of the $j$-th field.
We assume that each of these observations $\by_{j,\obs,\bc_{ij}}$ is equated with a noisy realization of the corresponding output field $\by_j$ of a multi-physics model $\M(\bx)$.
We represent a model evaluation of the $j$-th physical field with parameters $\bx$ at the $i$-th observed coordinate $\bc_{ij}$ with $\M_j(\bx,\bc_{ij})$.
% Introduce Gaussian likelihood
To compare model outputs and observations, we need a noise model that describes their statistical relationship.
This work assumes independent and identically distributed \new{(i.i.d.)} Gaussian noise with zero mean and a \new{known} diagonal covariance matrix.
We further assume that the field-specific noise $\noise_j$ is added to the model outputs $\M_j(\bx_\true, \bc_{ij})$ evaluated at the ground-truth model parameters $\bx_\true$ to obtain an observation $\by_{j, \obs, \bc_{ij}}$:
\begin{align}
    \by_{j, \obs, \bc_{ij}} = \M_j(\bx_\true, \bc_{ij}) + \noise_j \ \ \text{ with } \ \noise_j \sim \N(\bm{0}, \sigma_j^2 \bm{I}) \ .
    \label{eq:noise}
\end{align}
Our assumption implies a different covariance $\sigma_j^2 \bm{I}$ for each field $j$ and no correlation between the noise of observations or vector components.
Based on this noise assumption, we adopt a Gaussian multi-physics log-likelihood function:
\begin{align}
    \log p(\by_\obs|\M(\bx))
    \propto \sum^{N_{\fields}}_{j=1} \left( -\frac{1}{\sigma_j^2}  \sum^{N_{\obs,j}}_{i=1} \left\| \M_j(\bx, \bc_{ij}) - \by_{j, \obs, \bc_{ij}} \right\|^2 \right) \eeq
    \label{eq:log-likelihood}
\end{align}
This log-likelihood function clearly demonstrates the ease with which multi-physics observations are incorporated into Bayesian inverse analysis.

% Explain how we do BIA
The posterior distribution $p(\bx|\by_\obs)$ is generally not available in closed form.
Instead, only point-wise evaluations of the unnormalized posterior are possible.
These evaluations are obtained by computing the product of the prior $p(\bx)$ and the likelihood $p(\by_\obs|\M(\bx))$ at specific parameter values $\bx$.
In this work, which focuses on the multi-physics enhancement of Bayesian inverse analysis, we approximate the normalized posterior using a simple approach with two steps.
First, we evaluate the unnormalized posterior on a grid over the uncertain parameters $\bx$.
The grid is chosen to cover the high-density regions of the posterior.
This grid choice allows us to approximate the normalization constant (i.e., the denominator of Bayes' theorem given in Equation \ref{eq:bayes_rule}) in a second step using numerical integration with the trapezoidal rule.
We then divide the unnormalized posterior by the normalization constant to obtain an approximation of the posterior.
\new{To integrate a sufficiently smooth function numerically, the trapezoidal rule offers a convergence rate of $\mathcal{O}(1/n^2)$ in 1D and $\mathcal{O}(1/n)$ in 2D, where $n$ is the number of function evaluations.
However, the grid-based approach comes with the drawback that the high-density regions of the posterior must be known in advance to choose an appropriate grid.
If the high-density regions are not sufficiently resolved by the grid, the numerical error in the integration will be high.}
\delete{However}\new{Moreover}, the number of model evaluations using this grid-based approach grows exponentially with the number of uncertain parameters.
This inference approach is therefore feasible only for inverse problems with few uncertain parameters and a computationally inexpensive model $\M(\bx)$.
\delete{Moreover, the high-density regions of the posterior must be known in advance to choose an appropriate grid.}

Instead of grid-based approaches, two main classes of methods are commonly used to solve Bayesian inverse problems.
Although a thorough review of these methods is beyond the scope of this work, we provide a very brief summary here with entry points into the literature.
Clearly, all of the following methods can be combined with our approach of multi-physics enhancement.
The first class of methods for solving Bayesian inverse problems comprises \emph{sampling methods}.
These include \emph{Markov chain Monte Carlo (MCMC) methods} \cite{mosegaard2002, stuart2010, luengo2020}, which aim to generate a sequence of samples from the true posterior distribution.
Sampling methods also encompass \emph{particle methods} \cite{delmoral2006, chopin2020}, which approximate the posterior using a set of weighted samples, called particles in this context.
In the sequential case, these methods iteratively propagate particles from the prior to the posterior by repeatedly reweighting, resampling, and moving them.
The second class of methods for solving Bayesian inverse problems comprises \emph{variational methods} \cite{kingma2015, rezende2015, blei2017, nitzler2026}.
These methods approximate the posterior using a parameterized variational density.
The variational parameters of this density are optimized by minimizing a statistical distance measure (typically the Kullback-Leibler divergence) between the variational density and the true posterior.
Regardless of the solution method, \emph{surrogate models} \cite{kennedy2001, marzouk2007, ma2009, dinkel2024} can reduce the computational cost of repeatedly evaluating the model $\M(\bx)$ in the likelihood.

%%%%%%%%%%%%%%%%%%%%%%%%%%%%%%%%%%%%%%%%%%%%%%%%%%%%%%%%%%%%
\subsection{Coupling in multi-physics models}
\label{sec:coupled_multi-physics_models}

% Motivate why we need multi-physics models
\delete{Besides a multi-physics log-likelihood function, we also}\new{When moving from a single-physics to a multi-physics-enhanced Bayesian inverse analysis, we} need a computational model $\M$ that models each of the observed physical fields.
\new{We therefore extend the original single-physics model to a multi-physics model.}
Such a multi-physics model must account for the coupling between the fields.
Regarding inverse analysis, the type of coupling influences the information about the uncertain parameters that observations of each physical field can provide.
In this section, we take a closer look at the coupling types in multi-physics models with two fields and their potential to improve an inverse analysis.

We start with a single-physics model $\M_1(\bx) = \by_1$ with parameters $\bx$ and output field $\by_1$.
We then add a model $\M_2$ of a second physical field $\by_2$ such that observations of this second field can also provide information about the parameters $\bx$.
It is important to note that introducing the second physical field $\by_2$ is not primarily intended to improve forward predictions of the first field $\by_1$, although such improvements may occur.
Rather, the purpose of adding the second field is to enhance the inverse analysis of the parameters $\bx$.
This enhancement is possible if the second field $\by_2$ explicitly or implicitly depends on the parameters $\bx$.
An implicit dependency occurs when the second field $\by_2$ is coupled to the first field $\by_1$ since $\by_1$ explicitly depends on the parameters $\bx$.
We identify three different coupling types such that the second field $\by_2$ provides information about the parameters $\bx$.
These coupling types are presented in the following, with examples listed in Table \ref{tab:coupling_examples}.

%%%%%%%%%%%%
\begin{figure}[]
    \centering
    \vspace{-5pt}
    \includegraphics[scale=0.072]{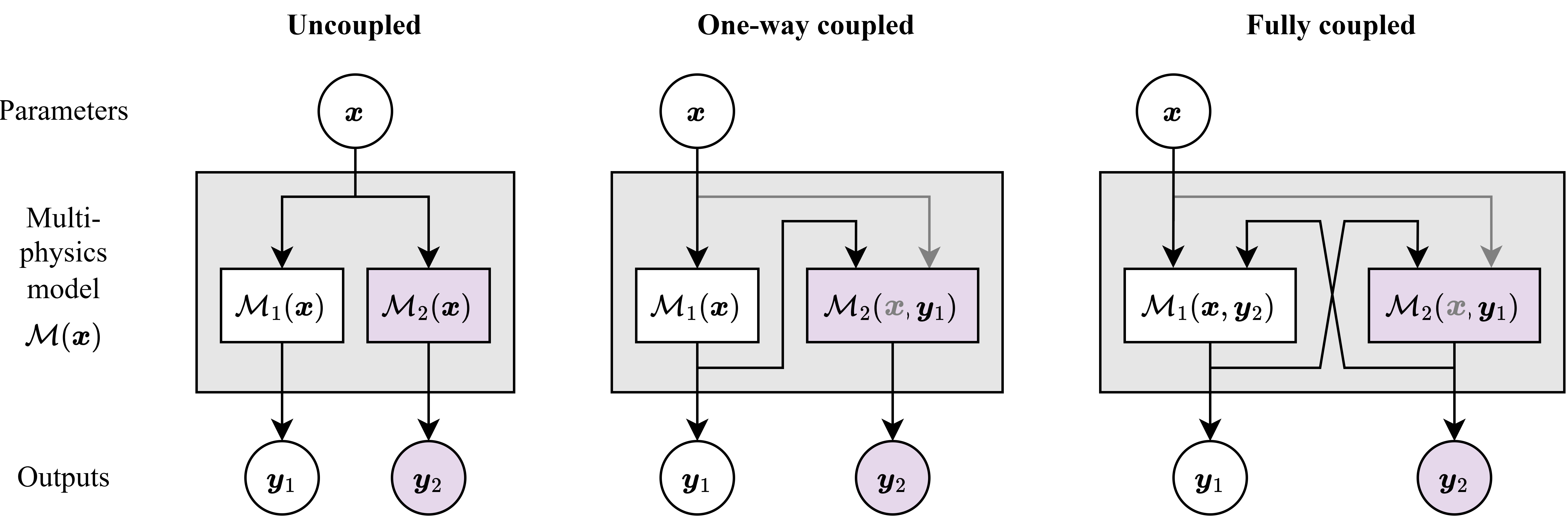}
    \caption{Coupling types of multi-physics models whose second output field $\by_2$ explicitly or implicitly depends on the model parameters $\bx$.
    Observations of such coupled second fields can enhance the inverse analysis of the model parameters $\bx$.
    The multi-physics model $\M(\bx)$ (\emph{gray box}) consists of the primary model $\M_1$ with output $\by_1$ and the secondary model $\M_2$ with output $\by_2$ (\emph{purple components}).
    Note that dependencies marked by arrows can be direct or through an integral or derivative.
    The explicit dependency of the secondary model $\M_2$ on the parameters $\bx$ is optional for one-way and fully coupled models as indicated by the grayed arrows.}
    \label{fig:coupling_types}
\end{figure}
%%%%%%%%%%%%

%%%%%%%%%%%%%%%%%%%%%%%%%%%%%
\begin{table}[]
\caption{Examples of multi-physics models with different coupling types. Observations of both the first and second output fields $\by_1$ and $\by_2$ of these models provide information about the parameters $\bx$.}
\label{tab:coupling_examples}
\centering
\begin{tabular}{ c c c } 
    \addlinespace
    \toprule
    \textbf{Uncoupled} & \textbf{One-Way Coupled} & \textbf{Fully Coupled} \\
    \midrule
    \makecell[l]{\greybf{Elastic deformation:} \\ $\bx$:\phantom{$_1$} Young's modulus \\ $\by_1$: Deflection under force \\ $\by_2$: Vibroacoustic signal}
    &
    \makecell[l]{\greybf{Fluid dynamics with smoke:} \\ $\bx$:\phantom{$_1$} Fluid viscosity
    \\ $\by_1$: Fluid velocity \\ $\by_2$: Smoke density}
    & 
    \makecell[l]{\greybf{Fluid-structure interaction:} \\ $\bx$:\phantom{$_1$} Fluid parameters \\ $\by_1$: Fluid velocity \\ $\by_2$: Structure deformation}
    \\ \addlinespace
    \makecell[l]{\greybf{Medical imaging:} \\ $\bx$:\phantom{$_1$} Density field \\ $\by_1$: Ultrasound imaging \\ $\by_2$: Computed tomography (CT) scan}
    & 
    \makecell[l]{\greybf{Heart contraction} \cite{corrado2015}\greybf{:} \\ $\bx$:\phantom{$_1$} Electrical state variables \\ $\by_1$: Electrocardiogram \\ $\by_2$: Myocardium displacement}
    & 
    \makecell[l]{\greybf{Natural convection:} \\ $\bx$:\phantom{$_1$} Thermal parameters \\ $\by_1$: Temperature field \\ $\by_2$: Fluid velocity}
    \\ \addlinespace
    \makecell[l]{\greybf{Marine electromagnetics} \cite{blatter2019}\greybf{:} \\ $\bx$:\phantom{$_1$} Seawater \& subsurface resistivities \\ $\by_1$: Magnetotellurics \\ $\by_2$: Controlled source electromagnetics}
    & 
    \makecell[l]{\greybf{Lung ventilation} \cite{dinkel2024}\greybf{:}\\ $\bx$:\phantom{$_1$} Lung stiffness field \\ $\by_1$: Air flow at ventilator \\ $\by_2$: EIT voltages }
    & 
    \makecell[l]{\greybf{Thermo-structure interaction} \cite{verdugo2016unified}\greybf{:} \\ $\bx$:\phantom{$_1$} Thermal stress field \\ $\by_1$: Structure deformation \\ $\by_2$: Temperature field } \\ 
    \bottomrule
\end{tabular}
\end{table}
%%%%%%%%%%%%%%%%%%%%%%%%%%%%%

% uncoupled
\paragraph{Uncoupled}
The left diagram in Figure \ref{fig:coupling_types} illustrates the first "coupling" type, in which both physical fields explicitly depend on the parameters $\bx$, but are not coupled to each other.
We call this coupling type \emph{uncoupled}.
This coupling type requires the least effort to enhance the inverse analysis of the parameters $\bx$.
No changes to the primary model $\M_1$ are necessary to add the second field, and an already existing model $\M_2$ of the second field can be reused.
In addition, the models $\M_1$ and $\M_2$ can be evaluated independently of each other.
% one-way coupled
\paragraph{One-way coupled}
The middle diagram in Figure \ref{fig:coupling_types} displays the second coupling type that allows us to use observations of the second field $\by_2$ to enhance the inverse analysis.
Here, the second field $\by_2$ depends on the first field $\by_1$ but not vice versa.
We refer to this coupling type as \emph{one-way coupled}.
An explicit dependency of the second field $\by_2$ on the parameters $\bx$ is optional for one-way coupled models.
This coupling type also requires little implementation effort since the primary model $\M_1$ remains unchanged, and a potentially available secondary model $\M_2$ can be reused.
Evaluation of a one-way coupled model is also straightforward, as we can evaluate the models $\M_1$ and $\M_2$ sequentially.
% fully coupled
\paragraph{Fully coupled}
The third coupling type that provides inverse information about the parameters $\bx$ in the second field $\by_2$ is presented in the right diagram of Figure \ref{fig:coupling_types}.
Here, both output fields $\by_1$ and $\by_2$ depend on each other, thus the name \emph{fully coupled}.
An explicit dependency of the second field $\by_2$ on the parameters $\bx$ is optional.
Implementing and evaluating fully coupled models is challenging due to the larger system matrices and interdependencies between fields.
Depending on the coupling strength, these models often require advanced numerical solution strategies to ensure stability and efficiency (see, e.g., \cite{verdugo2016unified}).

Information about the coupling type of a multi-physics model is provided by the model's system matrix.
We will briefly address this for multi-physics models with two output fields, although the following concepts also apply to models with an arbitrary number of physical fields \cite{verdugo2016unified}.
The physical fields of interest are typically governed by nonlinear partial differential equations that cannot be solved analytically. Instead, their solutions are approximated at discrete points in space and time. After discretization, a fully coupled model with two physical fields can be formulated as a set of nonlinear algebraic equations:
\begin{align}
  \begin{aligned}
    \f_1 (\by_1, \by_2) &= \bm{0} \ceq \\
    \f_2 (\by_1, \by_2) &= \bm{0} \eeq
    \label{eq:nonlinear-system}  
  \end{aligned}
\end{align}
The vector-valued functions $\f_j$ represent the discrete nonlinear equations associated with the $j$-th physical field, while the vectors $\by_j$ denote the corresponding discrete solutions.
We can solve the coupled nonlinear system of equations in \ref{eq:nonlinear-system} using a monolithic Newton method.
Starting from an initial guess $([\by_1, \by_2]^T)^{(0)}$, this method iteratively updates the solution vectors $([\by_1, \by_2]^T)^{(n)}$.
The superscript $(n)$ indicates quantities at the $n$-th Newton iteration.
In each of these iterations, the corrections $[\Delta \by_1, \Delta \by_2]^T$ are computed by linearizing equations $\f_1$ and $\f_2$ with respect to $\by_1$ and $\by_2$ and solving the resulting linear approximation
\begin{align}
    \underbrace{
    \begin{bmatrix}
        \partialfy{1}{1} & \partialfy{1}{2} \\[3ex]
        \partialfy{2}{1} & \partialfy{2}{2} 
    \end{bmatrix}
    ^{(n)}
    }_{\A^{(n)}}
    \begin{bmatrix}
        \Delta \by_1 \vphantompartial \\[3ex]
        \Delta \by_2 \vphantompartial
    \end{bmatrix}
    ^{(n+1)}
    = -
    \begin{bmatrix}
        \f_1 \vphantompartial \\[3ex]
        \f_2 \vphantompartial 
    \end{bmatrix}
    ^{(n)} .
    \label{eq:linearized-system}
\end{align}

The system matrix $\A^{(n)}$ helps us distinguish between the coupling types introduced in Figure \ref{fig:coupling_types} by identifying blocks in $\A^{(n)}$ that are zero regardless of the Newton iteration $n$:
\begin{align}
    \A &= 
    \underbrace{
        \begin{bmatrix}
            \dfrac{\partial \f_1}{\partial \by_1} & \bm{0} \\[3ex]
            \bm{0} & \dfrac{\partial \f_2}{\partial \by_2} 
        \end{bmatrix}
    }_{\text{uncoupled}} \ceq
    &
    \A &= 
    \underbrace{
        \begin{bmatrix}
            \dfrac{\partial \f_1}{\partial \by_1} & \bm{0} \\[3ex]
            \dfrac{\partial \f_2}{\partial \by_1} & \dfrac{\partial \f_2}{\partial \by_2} 
        \end{bmatrix}
    }_{\text{one-way coupled}} \ceq
    &
    \A &= 
    \underbrace{
        \begin{bmatrix}
            \dfrac{\partial \f_1}{\partial \by_1} & \dfrac{\partial \f_1}{\partial \by_2} \\[3ex]
            \dfrac{\partial \f_2}{\partial \by_1} & \dfrac{\partial \f_2}{\partial \by_2}
        \end{bmatrix}
    }_{\text{fully coupled}} \eeq
    \label{eq:system-matrices}
\end{align}
Note that the system matrices $\A^{(n)}$ are generally of high dimension.
Therefore, they are not inverted directly but with iterative solvers.
These solvers are most efficient if the system matrix $\A^{(n)}$ is sparse and well-conditioned.
As a result, solving fully coupled models generally requires more sophisticated solvers and more computational resources than solving uncoupled or one-way coupled models.

\FloatBarrier
%%%%%%%%%%%%%%%%%%%%%%%%%%%%%%%%%%%%%%%%%%%%%%%%%%%%%%%%%%%%
\section{Information gain in multi-physics-enhanced Bayesian inverse problems}
\label{sec:info_gain}

This section presents the information gain as a measure of the uncertainty reduction from the prior to the posterior distribution.
We also introduce the relative increase in information gain, which compares the uncertainty reduction achieved with single-physics versus multi-physics observations.

In this work, we measure the uncertainty reduction from the prior to the posterior distribution using the information gain.
This measure quantifies the amount of information gained about the parameters $\bx$ by including the observations $\by_\obs$.
The information gain is a well-established measure in information theory.
It is also widely used in Bayesian optimal experimental design to find an experimental design that maximizes the expected information gain \cite{chaloner1995bayesian, ryan2016review, foster2019variational, huan2024, orozco2024}.
In contrast to this use case, we use the information gain here to assess the additional reduction in uncertainty in a Bayesian inverse problem achieved by adding observations of a second physical field.

\begin{figure}[]
    \centering
    \begin{tikzpicture}
        \node[anchor=south west,inner sep=0] (img) at (0,0) {
            \includegraphics[scale=0.7]{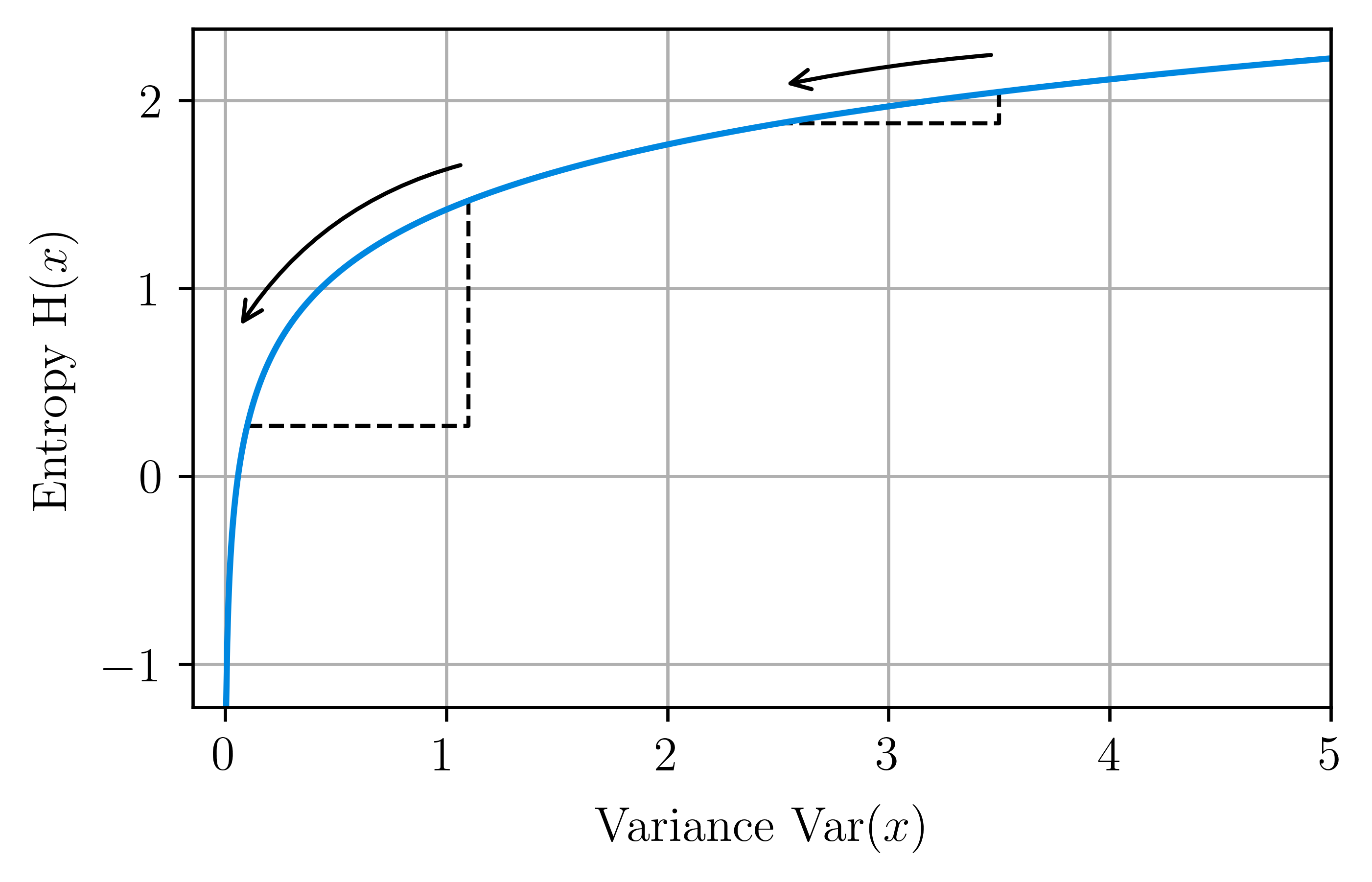}
        };

        \begin{scope}
            \path
            let
                \p1 = (img.north east)
            in
                node at ($(\x1*0.27, \y1*0.485)$) {\footnotesize 1.0}
                node at ($(\x1*0.38, \y1*0.64)$) {\footnotesize 1.20}
                node at ($(\x1*0.655, \y1*0.825)$) {\footnotesize 1.0}
                node at ($(\x1*0.77, \y1*0.868)$) {\footnotesize 0.17}
            ;
        \end{scope}
    \end{tikzpicture}
    \caption{
        Relationship between the variance $\Var{x}$ and the entropy $\Ent{x}=\frac{1}{2}\mylog\left(2 \pi e \Var{x}\right)$ of a univariate, normally distributed random variable $x$.
        For a fixed reduction in variance, the entropy decreases more when starting from a smaller variance.
        This property of the entropy reflects our intuition of uncertainty as the variance decreases.
    }
    \label{fig:entropy}
\end{figure}

\begin{figure}[]
\centering
\vspace{-5pt}
\includegraphics[width=0.95\textwidth]{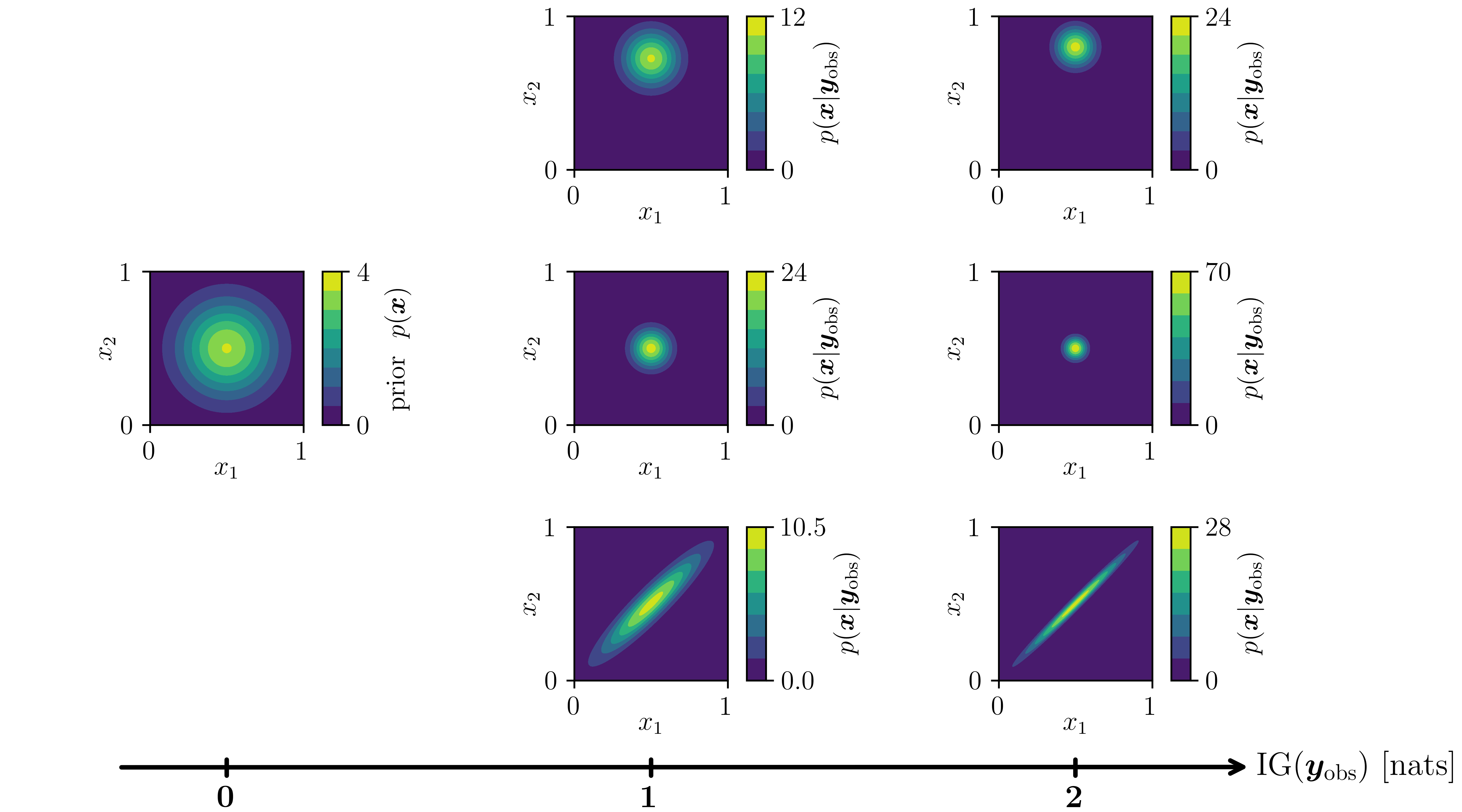}
\caption{Normally distributed posteriors $p(\bx | \by_\obs)$ with column-wise equal information gains $\IG{\by_\obs}$. The prior $p(\bx)$ of the two-dimensional random variable $\bx=\left[x_1, x_2\right]^T$ is depicted on the left.
The information gain originates from a shift in mean (\emph{upper row}), a reduction in diagonal covariance entries (\emph{upper and center row}), and an increase in off-diagonal covariance entries (\emph{lower row}).
}
\label{fig:axis_info_gain}
\end{figure}

% Introduce entropy and information gain
The information gain is based on the difference in (Shannon) entropy between the prior and the posterior distributions.
The entropy $\Ent{\bx}$ quantifies the expected information $\Inf{\bx}$ in a random variable $\bx$ with probability distribution $p(\bx)$ \cite{shannon1948mathematical}:
\begin{align}
    \Inf{\bx} = -\mylog\big(p(\bx)\big) \ \text{ with } \bx \sim p(\bx) \ceq
    \label{eq:information}
\end{align}
\vspace{-5mm}
\begin{align}
    \Ent{\bx} = \Exp{\Inf{\bx}} =-\int \mylog\big(p(\bx)\big)\cdot p(\bx) \dd \bx \eeq
    \label{eq:entropy}
\end{align}
This work relies on entropy instead of variance to measure uncertainty.
While the variance measures only the spread of a random variable around its mean, the entropy captures the uncertainty across the entire support of its probability distribution.
Moreover, the entropy reflects our intuition of uncertainty as the variance of a random variable with a single mode decreases.
If a parameter is highly uncertain, we expect its uncertainty to decrease slowly with decreasing variance.
However, if the uncertainty in the parameter is already low, we expect a stronger decrease in uncertainty as the variance decreases further.
The entropy exhibits this behavior for a univariate, normally distributed random variable, as shown in Figure \ref{fig:entropy}.
Based on these considerations, we rely on the entropy to compare the uncertainty in the prior and posterior distributions of an inverse problem.
In particular, we consider the information gain $\IG{\by_\obs}$, which is the relative entropy or Kullback-Leibler divergence between the prior distribution $p(\bx)$ and the posterior distribution $p(\bx|\by_\obs)$ \cite{kullback1951}:
\begin{equation}
    \IG{\by_\obs} = \Dkl{p(\bx|\by_\obs)}{p(\bx)} = \int \mylog\left(\frac{p(\bx|\by_\obs)}{p(\bx)}\right)\cdot p(\bx|\by_\obs)\dd \bx \eeq
    \label{eq:info_gain}
\end{equation}
To build intuition about information gain values, Figure \ref{fig:axis_info_gain} presents a selection of normally distributed posteriors and their associated information gains.

%%%%%%%%%%%%%%%%%%%%%%%%%%%%%%%%%%%%%%%
%%%%%%%%%%%%%%%%%%%%%%%%%%%%%%%%%%%%

% Introduce relative increase in information gain
The information gain $\IG{\by_\obs}$ quantifies how much the observations $\by_\obs$ reduce uncertainty when moving from the prior to the posterior distribution.
In this work, however, we are particularly interested in the additional uncertainty reduction from adding observations of a second physical field $\by_2$.
We quantify and compare this additional uncertainty reduction for varying numbers of observations, noise levels, and coupling types and strengths.
Our goal is to show that even a few or noisy observations of an additional field can considerably reduce uncertainty in the parameters $\bx$, even if this field is only weakly or one-way coupled.
To study this effect, we first compute the single-physics information gain $\IG{\by_{1,\obs}} = \Dkl{p(\bx|\by_{1,\obs})}{p(\bx)}$ using only the single-physics observations $\by_\obs = \by_{1,\obs}$.
Next, we compute the multi-physics information gain $\IG{\by_{1,\obs},\by_{2,\obs}} = \Dkl{p(\bx|\by_{1,\obs},\by_{2,\obs})}{p(\bx)}$ using multi-physics observations $\by_\obs = \left\{\by_{1,\obs}, \by_{2,\obs}\right\}$.
We then assess the effect of the second-field observations on the uncertainty reduction by computing the \emph{relative increase in information gain} $\RIIG{\by_{1,\obs}, \by_{2,\obs}}$:
\begin{align}
    \label{eq:riig}
    \RIIG{\by_{1,\obs}, \by_{2,\obs}}
    =
    \frac{\IG{\by_{1,\obs}, \by_{2,\obs}} - \IG{\by_{1,\obs}} }{ \IG{\by_{1,\obs}} } \eeq
\end{align}

\begin{figure}[]
    \centering
    \vspace{-5pt}
    \begin{tikzpicture}
        \node[anchor=south west,inner sep=0] (img) at (0,0) {
            \includegraphics[width=0.95\textwidth]{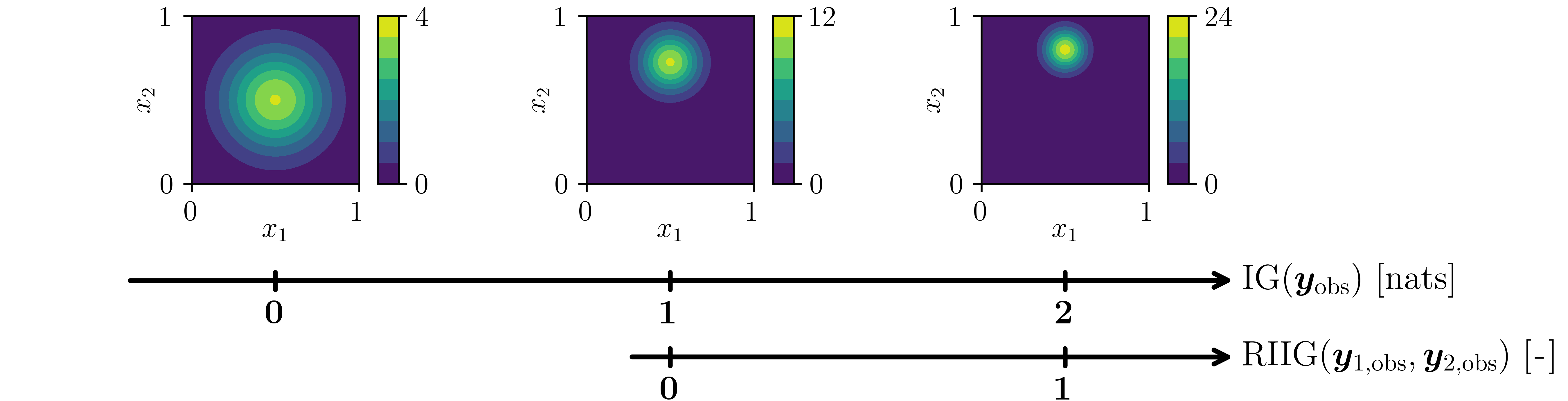}
        };

        \begin{scope}
            \path
            let
                \p1 = (img.north east)
            in
                node at ($(\x1*0.175, \y1*1.15)$) {\small Prior}
                node at ($(\x1*0.175, \y1*1.05)$) {\small $p(\bx)$}
                node at ($(\x1*0.435, \y1*1.15)$) {\small Single-physics posterior}
                node at ($(\x1*0.435, \y1*1.05)$) {\small $p(\bx|\by_{1,\obs})$}
                node at ($(\x1*0.685, \y1*1.15)$) {\small Multi-physics posterior}
                node at ($(\x1*0.685, \y1*1.05)$) {\small $p(\bx|\by_{1,\obs},\by_{2,\obs})$}
            ;
        \end{scope}
    \end{tikzpicture}
    \caption{
        Exemplary prior and posteriors of a Bayesian inverse analysis with single-physics observations $\by_{\obs} = \by_{1,\obs}$ and multi-physics observations $\by_{\obs} = \{\by_{1,\obs}, \by_{2,\obs}\}$.
        The information gain $\IG{\by_{\obs}}$ quantifies the uncertainty reduction from the prior to the posterior.
        The relative increase in information gain $\RIIG{\by_{1,\obs}, \by_{2,\obs}}$ quantifies the additional uncertainty reduction achieved by adding observations $\by_{2,\obs}$ of the second physical field to the inverse analysis (see Equation \ref{eq:riig}).
    }
    \label{fig:ig_and_riig}
\end{figure}

Figure \ref{fig:ig_and_riig} compares the information gain and the relative increase in information gain for an exemplary prior, single-physics posterior, and multi-physics posterior.

In this work, we assume that the additional physical fields do not introduce additional uncertain model parameters.
If this assumption does not hold, any additional uncertain parameters must be marginalized, which would likely reduce the increase in information gain from adding the second-field observations.

\FloatBarrier

%%%%%%%%%%%%%%%%%%%%%%%%%%%%%%%%%%%%%%%%%%%%%%%%%%%%%%%%%%%%
%%%%%%%%%%%%%%%%%%%%%%%%%%%%%%%%%%%%%%%%%%%%%%%%%%%%%%%%%%%%
\section{Numerical demonstration}

% Goal of the numerical demonstration
In this section, we demonstrate the potential of multi-physics observations and modeling to reduce uncertainty in the posterior distribution of an \new{orginally single-physics} Bayesian inverse analysis.
% Key aspects being tested
In particular, we examine the relative increase in information gain (RIIG) when moving from single-physics to multi-physics observations.
% Description of the test cases
Section \ref{sec:simple_model} presents a Bayesian inverse problem based on a scalar-valued, one-way coupled electromechanical model.
Using this model, we compute the RIIG for varying numbers of observations and noise levels of the second physical field.
In Section \ref{sec:poro_model}, we use a fully-coupled mixed-dimensional multiphase porous medium model to solve a Bayesian inverse problem.
Here, we vary the noise levels in the first and second observed fields as well as the coupling strength to assess the impact on the RIIG.
% Additional information / Disclaimer
These inverse problems and models are not claimed to be relevant real-world applications.
They are purely selected to showcase the potential of the multi-physics-enhanced approach to reduce uncertainty in the posterior distribution.

To solve the inverse problems, we used synthetic observations that are noisy realizations of the ground-truth model output according to Equation \ref{eq:noise}.
Instead of prescribing the variance $\sigma_j^2$ of the Gaussian noise $\noise_j$, we prescribed the signal-to-noise ratio (SNR) of the ground-truth model output $\M_j(\bx_\true,\bc_{ij})$ for each field $j$:
\begin{align}
    \SNR_j = \frac{\Exp{\|\M_j(\bx_\true,\bc_{ij})\|^2}}{\Exp{\|\noise_j\|^2}}  = \frac{ \frac{1}{N_{\obs, j}} \sum^{N_{\obs, j}}_{i=1}{\|\M_j(\bx_\true,\bc_{ij})\|^2}}{\dim(\M_j(\bx_\true,\bc_{ij})) \cdot \sigma_j^2} \eeq
    \label{eq:snr}
\end{align}
The SNR is a dimensionless quantity that facilitates the comparison of noise levels across physical fields.
We derived the field-specific noise variance $\sigma_j^2$ for a prescribed $\SNR_j$ by rearranging Equation \ref{eq:snr}.

All code and scripts used for the numerical demonstrations are publicly available at \href{https://github.com/leahaeusel/mpebia}{https://github.com/leahaeusel/mpebia}.

\FloatBarrier
%%%%%%%%%%%%%%%%%%%%%%%%%%%%%%%%%%%%%%%%%%%%%%%%%%%%%%%%%%%%
\subsection{Multi-physics-enhanced Bayesian inverse analysis with a scalar-valued one-way coupled electromechanical model}
\label{sec:simple_model}

% Purpose of the Demonstration / Why do we choose this model?
In this simple demonstration example, the task is to infer the Young's modulus $E$ and Poisson's ratio $\nu$ of a material.
For this Bayesian inverse problem, the vector of uncertain model parameters is $\bx = [E, \nu]^T$.
We start with a mechanical single-physics model and determine the posterior of the parameters after observing displacements from a tensile test.
We then demonstrate the improvement in the posterior's information content by adding a second, one-way coupled physical field as an additional source of information.
This second field consists of electrical measurements, assuming the material of interest is electrically conductive.
The electrical field
\footnote{By \emph{electrical field}, we refer here to the general physical field of electricity, encompassing quantities such as voltage, electric current, and electrical resistance.
In contrast, \emph{electric field} denotes the vector field that describes the force per unit charge.
In the following, we refer explicitly to the general electrical field, not the electric field.
}
depends on the mechanical field through the displacement, which affects the electrical resistance.
For simplicity, we assume that the electrical field does not influence the mechanical field.
This one-way coupled electromechanical model was inspired by \cite{roth2015}.

\subsubsection{Forward model}

\begin{figure}[]
    \centering
    \hspace{-5pt}
    \includegraphics[scale=0.075]{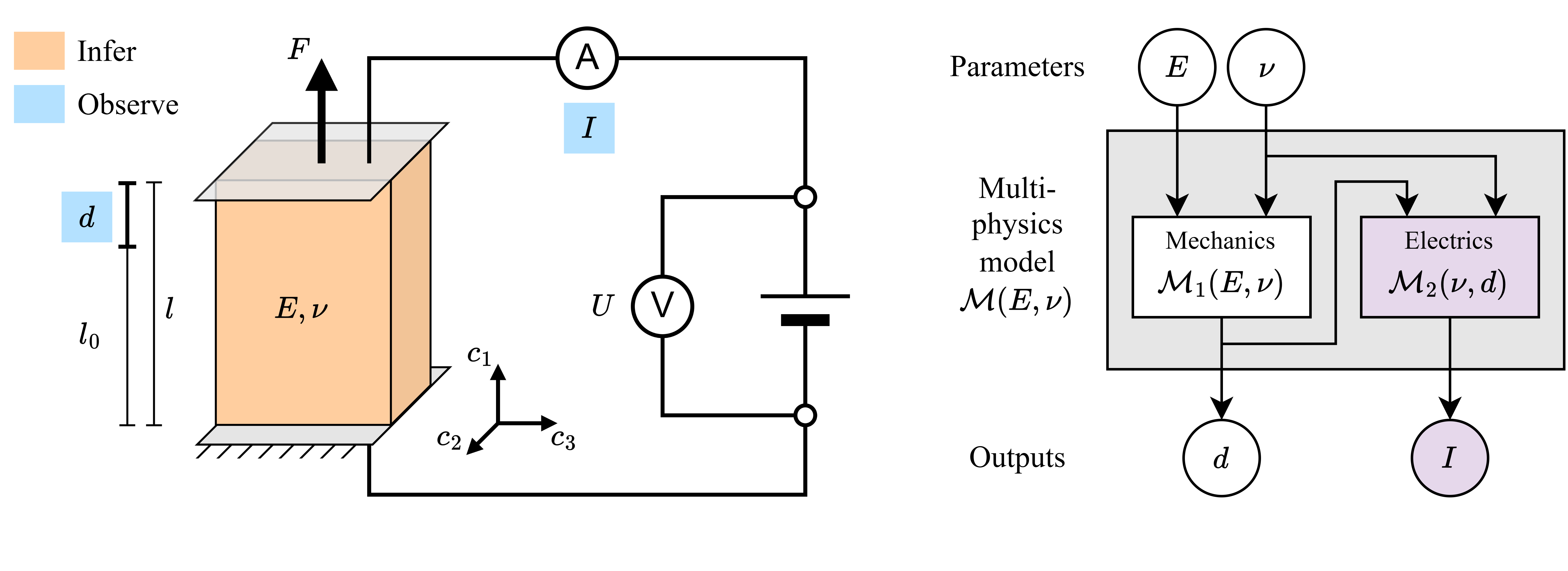}
    \caption{\emph{Left:} Tensile test of a cube with concurrent electric measurements. We observe the displacement $d$ of the cube's upper surface and the electric current $I$ in dependence on the applied force $F$. The Young's modulus $E$ and the Poisson's ratio $\nu$ of the cube are to be inferred by Bayesian inverse analysis.
    \emph{Right:} The resulting one-way coupled electromechanical model $\M(E, \nu)$. The mechanical model $\M_1$ is independent of the electrical model $\M_2$. The electrical model $\M_2$ depends on the mechanical model $\M_1$ through the displacement $d$.
    Purple components are only added to the model to enhance the inverse analysis of $E$ and $\nu$, not the forward prediction of $d$.}
    \label{fig:simple_model}
\end{figure}

\paragraph{The first, mechanical field}
Our model considers a cube made of the material of interest.
The material is isotropic and homogeneous and is modelled by a Saint Venant-Kirchhoff material model.
The cube has side lengths $l_0 = \SI{10}{mm}$ and thus a square cross-sectional area $A_0=\left(l_0\right)^2$ in its undeformed state.
Starting from this state, a force $F$ acts on the upper surface of the cube in $c_1$-direction while the cube's lower surface is fixed in $c_1$-direction.
As a result, the cube elongates to length $l=l_0 + d$ with $d$ being the displacement of the cube's upper surface.
The lateral surfaces of the cube normal to the $c_2$-direction are free to move, while the lateral surfaces normal to the $c_3$-direction are fixed in the $c_3$-direction.
Figure \ref{fig:simple_model} displays this setup.
To infer the parameters of interest $E$ and $\nu$, we observe the displacement $d$ depending on the applied force $F$.
We derive the relation between force and displacement by assuming that the cross-sectional area of the cube stays square and normal to $c_1$ during the deformation.
We further neglect any gravitational effects.
The resulting mechanical model is described by
\begin{align}
    f_1(d) = 2 l_0^2 d + 3 l_0 d^2 + d^3 - \frac{2 F l_0}{E} \left(1 - \nu ^2\right) = 0 \eeq
    \label{eq:mechanical_problem}
\end{align}
The derivation of Equation \ref{eq:mechanical_problem} can be found in Appendix \ref{ap:simple_model}.

%%%%%%%%%%%%
%%%%%%%%%%%%

% Electrical model
\paragraph{The second, electrical field}
To obtain more information about the parameters of interest $E$ and $\nu$, we add a second physical field to the tensile test in the form of electrical measurements.
Figure \ref{fig:simple_model} shows that a voltage $U = \SI{10}{V}$ is applied between the upper and lower surfaces of the cube.
The electric potential is constant across each of these surfaces.
We observe the electric current $I$ resulting from the potential difference.
The electric current depends on the resistance of the cube, which in turn depends on the displacement $d$ and the deformed cross-sectional area of the cube.
Due to this dependency, observations of the electric current $I$ contain information about the mechanical problem.
Appendix \ref{ap:simple_model} describes the derivation of the resulting electrical model:
\begin{align}
    f_2(d, I) = \rho (l_0 + d) I - U l_0 \sqrt{ l_0^2 - \frac{\nu}{1-\nu} (2 l_0 d + d^2)} = 0 \eeq
    \label{eq:electrical_problem}
\end{align}
We assume that the electrical resistivity $\rho = \SI{1.0}{\Omega m}$ of the material of interest is known and constant for all measurements.

% Describe coupling
\paragraph{Coupling between the fields}
For given parameters $E$ and $\nu$, we can find variables $d$ and $I$ that satisfy the coupled, nonlinear Equations \ref{eq:mechanical_problem} and \ref{eq:electrical_problem}.
We describe such an evaluation of the multi-physics model by $[d, I]^T = \M(E,\nu)$. 
Figure \ref{fig:simple_model} illustrates the partition of this multi-physics model $\M$ into the mechanical model $d = \M_1(E, \nu)$ and the electrical model $I = \M_2(\nu, d)$.
The electrical model $\M_2$ directly depends on the parameter of interest $\nu$ but is also coupled to the mechanical model through the displacement $d$.
By contrast, the mechanical model $\M_1$ is independent of the electrical field.
Thus, the presented electromechanical multi-physics model $\M$ is one-way coupled.

\paragraph{Linearized system of equations}
To evaluate this one-way coupled multi-physics model, we solve the nonlinear Equations \ref{eq:mechanical_problem} and \ref{eq:electrical_problem} with a monolithic Newton method.
Hence, we linearize the two equations as previously shown in Equation \ref{eq:linearized-system}:
\renewcommand{\arraystretch}{2.0}
\begin{align}
    \begin{bmatrix}
        2 l_0^2 + 6 l_0 d + 3 d^2 & 0\\
        \rho I - \frac{U l_0 (1+2d)}{2 \sqrt{(2 l_0 d + d^2) \frac{\nu}{\nu-1} + l_0^2}} \frac{\nu}{\nu -1}  & \rho (l_0 + d)
    \end{bmatrix}
    ^{(n)}
    \begin{bmatrix}
        \Delta d \vphantom{2 l_0^2}\\ 
        \Delta I \vphantom{\frac{U l_0 (1+2d)}{2 \sqrt{(2 l_0 d + d^2) \frac{\nu}{\nu-1}  + l_0^2}}}
    \end{bmatrix}
    ^{(n+1)}
    = -
    \begin{bmatrix}
        2 l_0^2 d + 3 l_0 d^2 + d^3 - \frac{2 F l_0}{E} \left(1 - \nu ^2\right) \\ 
        \rho (l_0 + d) I - U l_0 \sqrt{(2 l_0 d + d^2) \frac{\nu}{\nu-1} + l_0^2}
    \end{bmatrix}
    ^{(n)} \eeq
    \label{eq:linearized_simple_model}
\end{align}
In the notation of Section \ref{sec:coupled_multi-physics_models}, the solution vectors $\by_1$ and $\by_2$ of the two fields are both one-dimensional.
They correspond to the displacement $y_1 = d$ for the first, mechanical field and the electric current $y_2 = I$ for the second, electrical field.
Note that the upper-right system matrix entry corresponding to $\frac{\partial f_1}{\partial y_2}$ in Equation \ref{eq:linearized_simple_model} is zero because the model is one-way coupled.

% INVERSE METHODS

% Observations, noise, and prior
\subsubsection{Observations}
\label{sec:model1_observations}
We generated the synthetic observations based on evaluations of the electromechanical model with the ground-truth parameters $E_\true = \SI{11}{kPa}$ and $\nu_\true = 0.35$ at varying force values $F$.
For a given number of observations $N_{\obs, j}$ of the $j$-th physical field, these force values were equidistant to each other and the minimal and maximal force values of $F_\text{min} = \SI{0.0}{N}$ and $F_\text{max} = \SI{0.4}{N}$.
To obtain the observations, we added zero-mean Gaussian noise to the model outputs (compare Equation \ref{eq:noise}).
The variance of the noise $\sigma_j^2$ was computed from the signal-to-noise ratio $\SNR_j$ of each field $j$ according to Equation \ref{eq:snr}.
We generated the noise samples using a Sobol' sequence \cite{sobol1967} to ensure they were evenly distributed and representative of the underlying normal distribution, even for a small number of samples.
For the mechanical field, we always used the same displacement observations with a relatively low $\SNR_1 = 50$ and $N_{\obs, 1} = 16$ observations.
For the electrical field, we varied the $\SNR_2$ and the number of observations $N_{\obs, 2}$ to evaluate the relative increase in information gain for different configurations
\footnote{
As mentioned above, this inverse example is not intended to represent any real, practical application.
However, the multi-physics-enhanced BIA in this example can be motivated by the unavailability of cross-sectional measurements during deformation.
Only top and bottom measurements are accessible.
}.

\subsubsection{Probabilistic model \& inference method}
For the prior distribution of the parameters of interest $E$ and $\nu$, we chose a truncated normal distribution according to Table \ref{tab:params_model1}.
The truncation bounds $\ba$ and $\bb$ ensure that physically implausible values of $E$ and $\nu$ have a probability density of zero.
\new{Assuming i.i.d.\! Gaussian noise with known variances, we}\delete{We further} used the Gaussian multi-physics log-likelihood function introduced in Equation \ref{eq:log-likelihood}.
The likelihood variance $\sigma_j^2$ of each field $j$ was always identical to the variance used to sample the observational noise as described in Section \ref{sec:model1_observations}.
With this prior and likelihood, we evaluated the posterior on a $100\times100$ grid of $E$ and $\nu$.
The grid points in each dimension were evenly spaced with respect to the cumulative distribution function of the prior.
\new{We also repeated the following analyses with finer grids of $E$ and $\nu$ to confirm the overall trends.}

\renewcommand{\arraystretch}{1.5}
\begin{table}[b]
    \caption{
        Parameters of the prior used for the Bayesian inverse analyses using the electromechanical model.
    }
	\label{tab:params_model1}
	\centering
	\begin{tabular}{lll}
        \addlinespace
		\toprule
		Parameter &
        Notation & 
        Value or Distribution \\
        \midrule
        Prior distribution & $p( \bx = [E, \nu]^T )$ & $\mathcal{TN} \left( \bmu,\Sigma,\ba,\bb \right) $ \\
        Prior mean & $\bmu$ & $\left[\SI{10}{kPa}, 0.3\right]^T$ \\
        Prior covariance & $\Sigma$ & $\diag\left((\SI{2}{kPa})^2, (0.15)^2\right)$ \\
        Lower truncation bound & $\ba$ & $\left[\SI{0}{kPa}, 0\right]^T$ \\
        Upper truncation bound & $\bb$ & $\left[\infty, 0.5\right]^T$ \\
		\bottomrule
	\end{tabular}
\end{table}

\subsubsection{Comparison of a single-physics and two multi-physics-enhanced Bayesian inverse analyses}
\label{sec:model1_posteriors}

\paragraph{Method}
% Method 1st figure
First, we compare the observations and results of a single-physics Bayesian inverse analysis (BIA) with those of two multi-physics-enhanced BIAs.
While the single-physics BIA only considers observations of the displacement $d$, the multi-physics-enhanced BIAs use observations of both the displacement $d$ and the electric current $I$.
For the first multi-physics-enhanced BIA, we use a small number of electric current observations ($N_{\obs, 2}=2$) with a relatively high SNR ($\SNR_2=1.2\times10^4$, low noise).
The second multi-physics-enhanced BIA uses a large number of electric current observations ($N_{\obs, 2}=256$) with a relatively low SNR ($\SNR_2=80$, high noise).

%%%%%%%%%%%%%
\begin{figure}[p]
    \centering
    \includegraphics[width=\textwidth]{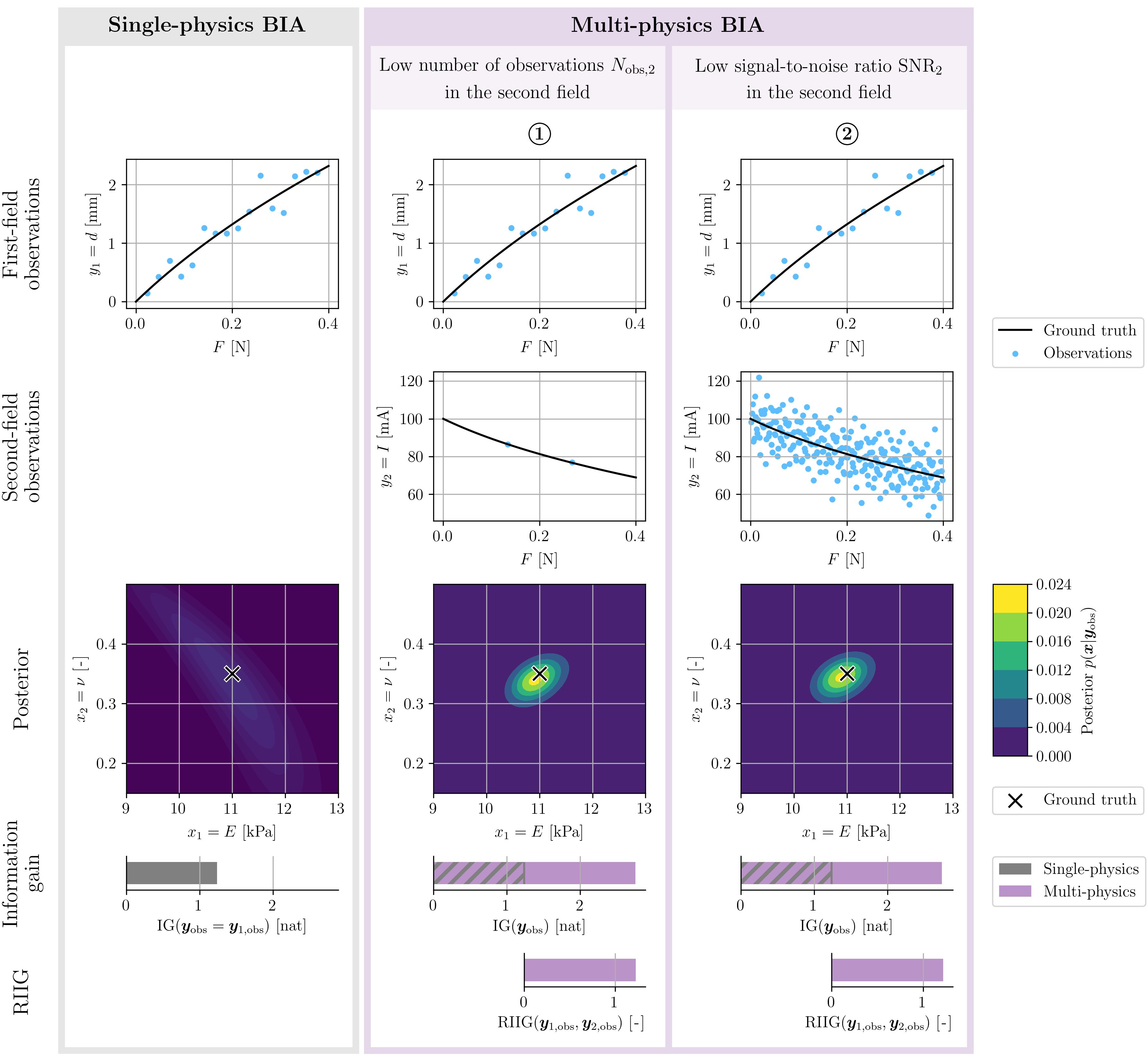}
    \caption{Observations and results of single-physics (\emph{left}) versus multi-physics-enhanced (\emph{middle and right}) Bayesian inverse analysis (BIA) using the one-way coupled electromechanical model.
    Even though the second-field observations are low in number (\emph{middle}) or high in noise (\emph{right}), both multi-physics posteriors are considerably more concentrated and have a higher information gain than the single-physics posterior.
    The relative increase in information gain of the two multi-physics-enhanced BIAs is very similar, with $\RIIG{\by_{1, \obs}, \by_{2, \obs}} = 1.23$ (\emph{middle}) and $1.22$ (\emph{right}) although their observations of the second, electrical field differ strongly in signal-to-noise ratio and number.
    }
    \label{fig:model1_posteriors}
\end{figure}
%%%%%%%%%%%%%

%%%%%%%%
\begin{figure}[]
    \vspace{-5pt}
    \centering
    \includegraphics[width=\textwidth]{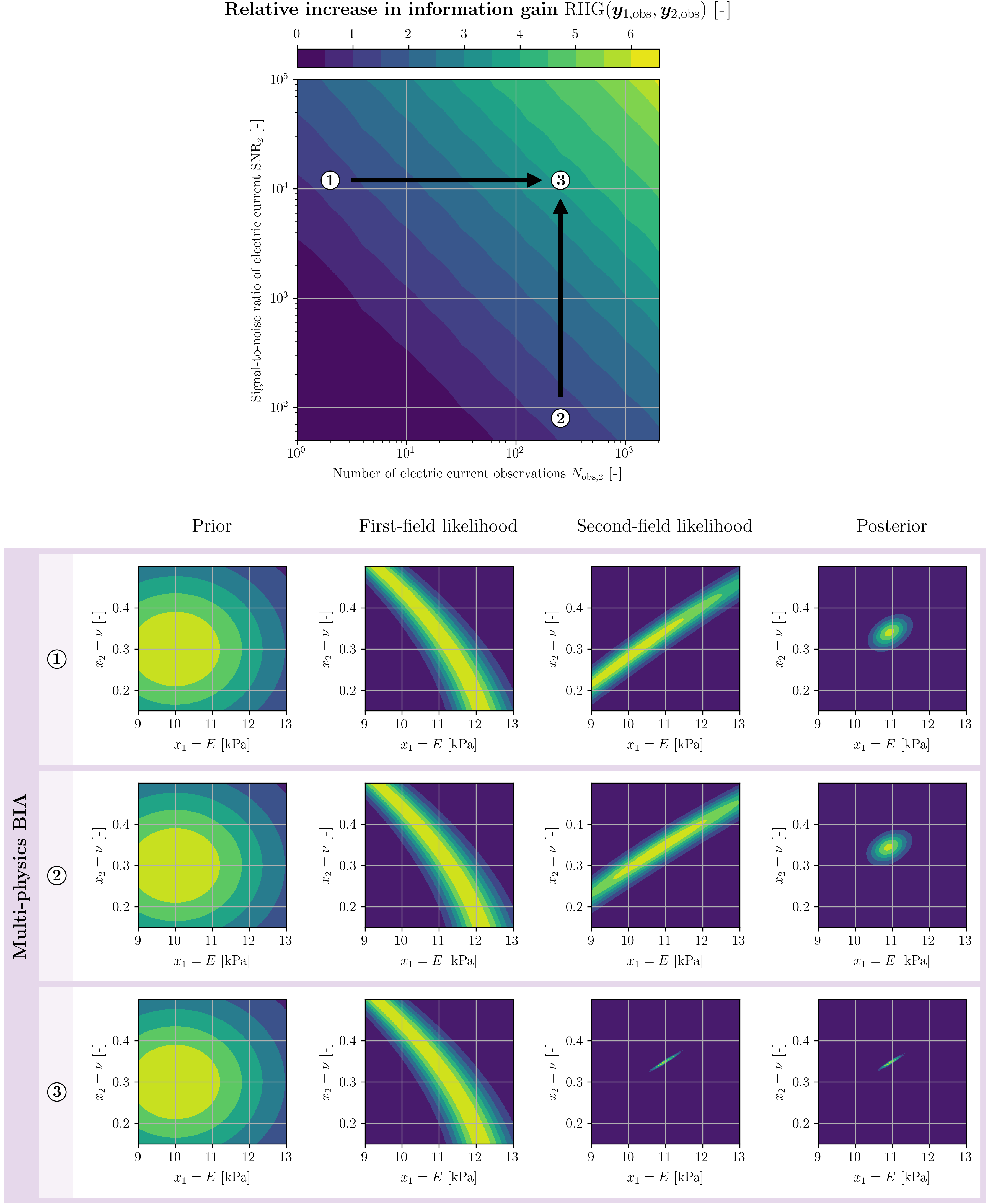}
    \vspace{-5pt}
    \caption{
        \delete{\emph{Left:}}\new{\emph{Top:}} The relative increase in information gain (RIIG) becomes larger for a higher number of observations $N_{\obs, 2}$ and a higher signal-to-noise ratio $\SNR_2$ in the second, electrical field.
        Note that both axes are scaled logarithmically.
        \delete{\emph{Right:} The multi-physics posteriors for second-field observations with $N_{\obs, 2}$ and $\SNR_2$ and as indicated by the circled numbers in the left plot.
        Having a higher number of observations $N_{\obs, 2}$ and a higher $\SNR_2$ in the second, electrical field results in a more concentrated posterior.}
        \new{\emph{Bottom:} The prior $p(\bx)$, the likelihoods of each observed field $p(\by_{1, \obs}|\bx)$ and $p(\by_{2, \obs}|\bx)$, and the posterior $p(\bx|\by_{1, \obs}, \by_{2, \obs})$ for three different multi-physics Bayesian inverse analyses (BIAs). The three BIAs differ in $N_{\obs, 2}$ and $\SNR_2$ of the second-field observations as indicated by the circled numbers in the top plot.
        The likelihood of the second field increasingly constrains the posterior for higher $N_{\obs, 2}$ and $\SNR_2$.
        }
    }
    \label{fig:model1_riig}
\end{figure}
%%%%%%%%

\paragraph{Results \& discussion}
% Results and Discussion 1st figure
Figure \ref{fig:model1_posteriors} shows the observations, posteriors, and information gains of the single-physics BIA versus the two multi-physics-enhanced BIAs.
The single-physics BIA results in an uninformative, widespread posterior of the parameters $E$ and $\nu$.
The first multi-physics posterior is considerably more concentrated, even though we used only two additional electric current observations.
The first multi-physics information gain is also considerably higher than the single-physics information gain, resulting in a relative increase in information gain of $\RIIG{\by_{1, \obs}, \by_{2, \obs}}=1.23$.
This considerable improvement, even with a few additional observations, can be explained by the high SNR of the electric current observations in the first multi-physics-enhanced BIA.
The second multi-physics-enhanced BIA shows a similar improvement in the posterior and the information gain, although the electric current observations have a low SNR.
The relative increase in information gain is also almost identical to that of the first multi-physics posterior with $\RIIG{\by_{1, \obs}, \by_{2, \obs}}=1.22$.
In this case, the large number of observations in the second field mitigates the low SNR.
These two multi-physics-enhanced BIAs demonstrate that even a few or noisy observations of an additional one-way coupled field can considerably reduce uncertainty in the posterior.
It should be noted that the one-way coupling allows for a relatively easy implementation and evaluation of the additional field, especially if a readily available model can be used.

\subsubsection{Relative increase in information gain for varying second-field observations}

\paragraph{Method}
% Method 2nd figure
Having examined two individual multi-physics posteriors and their \new{respective (relative increase) in} information gain\delete{s} in Section \ref{sec:model1_posteriors}, we now investigate the relative increase in information gain (RIIG) for a range of second-field observations.
For this investigation, we varied the number $N_{\obs, 2}$ and $\SNR_2$ of the electric current observations and computed the resulting RIIG as introduced in Equation \ref{eq:riig}.
We evaluated the RIIG on a $50 \times 12$ log-spaced grid of $N_{\obs, 2}$ and $\SNR_2$.
The displacement observations were constant throughout the investigation, as already shown in Figure \ref{fig:model1_posteriors}.

\paragraph{Results \& discussion}
% Results 2nd figure
The \delete{left}\new{top} plot in Figure \ref{fig:model1_riig} shows the RIIG increasing with the number of observations and the SNR of the second-field observations.
Each point in the plot corresponds to a multi-physics-enhanced BIA.
This is also illustrated by the \delete{right}\new{bottom} plot\new{s} in Figure \ref{fig:model1_riig}, which show\delete{s} the \new{prior, the likelihoods of each observed field, and the }multi-physics posteriors at specific points in the \delete{left}\new{top} plot.
The posteriors 1 and 2 were already presented in Figure \ref{fig:model1_posteriors}.
\new{Although their likelihoods of the two observed fields are widely spread individually, they intersect only in a small region of the parameter space, leading to concentrated posteriors and an RIIG of 1.23 and 1.22, respectively.}
Combining the high $\SNR_2$ of posterior 1 with the large number of observations $N_{\obs, 2}$ of posterior 2 results in posterior 3.
This posterior is highly concentrated and has a RIIG of $\RIIG{\by_{1, \obs}, \by_{2, \obs}}=3.65$.
Our results demonstrate that observations of a one-way coupled field can substantially reduce the uncertainty in the posterior compared to using single-physics observations alone.
This reduction is particularly pronounced when the second-field observations have a high SNR and a high number of observations.
However, it should be stressed again that even a few second-field observations can considerably impact the uncertainty if their SNR is sufficiently high.
The same holds for observations with a low SNR if a sufficient number of observations is available.

\FloatBarrier
%%%%%%%%%%%%%%%%%%%%%%%%%%%%%%%%%%%%%%%%%%%%%%%%%%%%%%%%%%%%%%%%%%%
\subsection{Multi-physics-enhanced Bayesian inverse analysis with a complex, fully coupled porous medium model}
\label{sec:poro_model}

The goal of the Bayesian inverse analysis in this section is to infer the Young's moduli $E^h$ and $E^d$ in a healthy and a diseased region of a lung tissue patch.
For this inverse problem, the vector of uncertain model parameters is $\bx = [E^h, E^d]^T$.
We first determine the posterior of these parameters after observing the displacement of the lung tissue $\bm{d}^t$ at end-inspiration.
Following the multi-physics-enhanced approach for Bayesian inverse analysis, we then also incorporate observations of the blood volume fraction $\varepsilon^b$ in the lung tissue.

%%%%%%%%%%%%%%%%%%%%%%%
\begin{figure}[]
    \centering
    \begin{tikzpicture}
        \node[anchor=south west,inner sep=0] (img) at (0,0) 
            {\includegraphics[width=0.31\textwidth]{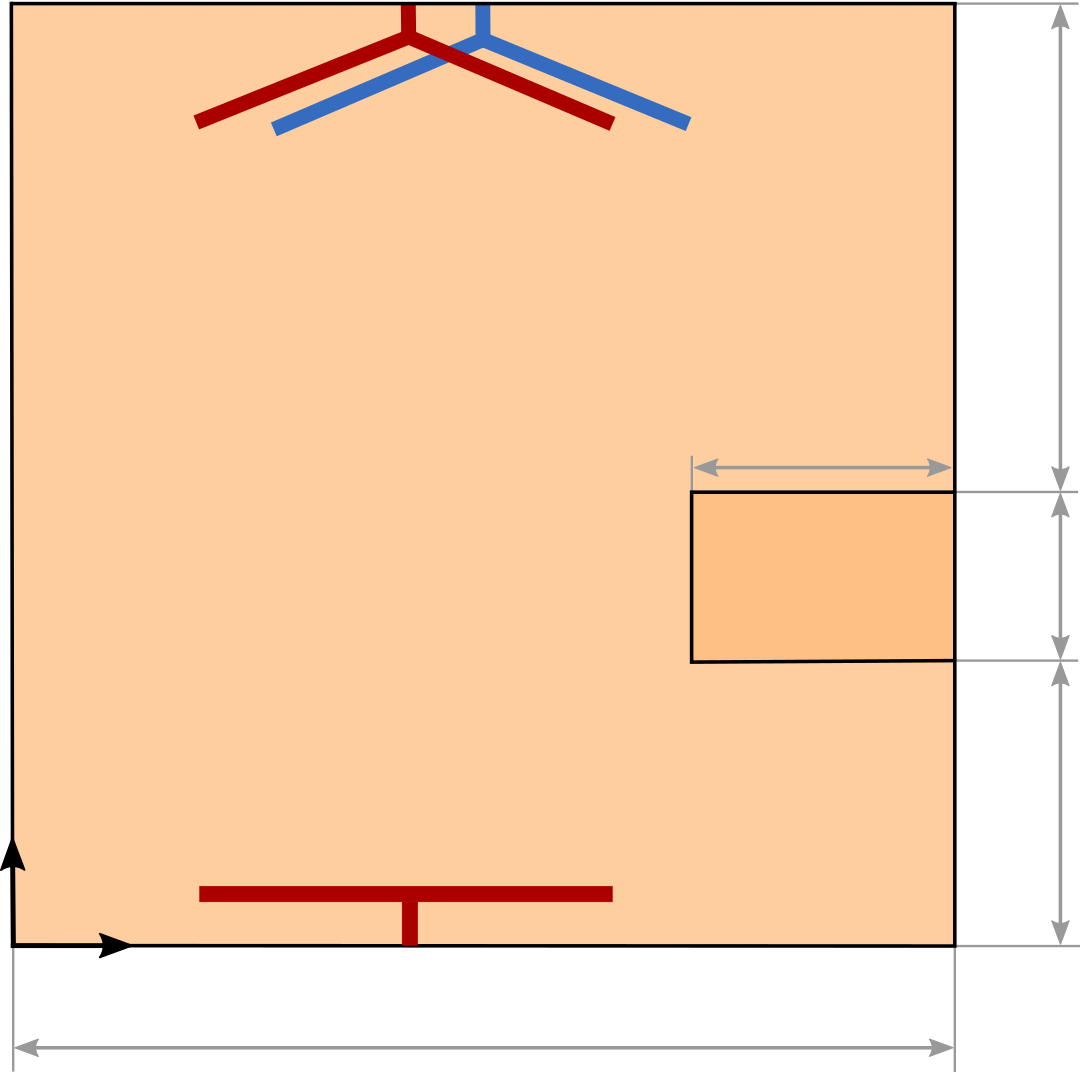}};

        \begin{scope}
            \definecolor{blue}{RGB}{53,108,192}  
            \definecolor{red}{RGB}{170,0,0}  
            \definecolor{grey}{RGB}{153,153,153}  
            \path
            let
                \p1 = (img.north east)
            in
                node at ($(\x1*0.11, \y1*0.075)$) {$c_1$}
                node at ($(\x1*-0.035, \y1*0.2)$) {$c_2$}
                node at ($(\x1*0.4, \y1*0.55)$) {\large $E^h$}
                node at ($(\x1*0.76, \y1*0.46)$) {\large $E^d$}
                node[blue] at ($(\x1*0.7, \y1*0.95)$) {\footnotesize Airway}
                node[red] at ($(\x1*0.15, \y1*0.95)$) {\footnotesize Artery}
                node[red] at ($(\x1*0.375, \y1*0.22)$) {\footnotesize Vein}
                node[grey] at ($(\x1*0.77, \y1*0.6)$) {\footnotesize \SI{2.5}{mm}}
                node[grey] at ($(\x1*0.45, \y1*-0.02)$) {\footnotesize \SI{10}{mm}}
                node[grey, rotate=90] at ($(\x1*1.03, \y1*0.25)$) {\footnotesize \SI{3}{mm}}
                node[grey, rotate=90] at ($(\x1*1.03, \y1*0.46)$) {\footnotesize \SI{2}{mm}}
                node[grey, rotate=90] at ($(\x1*1.03, \y1*0.75)$) {\footnotesize \SI{5}{mm}}
            ;
        \end{scope}
    \end{tikzpicture}
    \hspace{1.1cm}
    \includegraphics[scale=0.07]{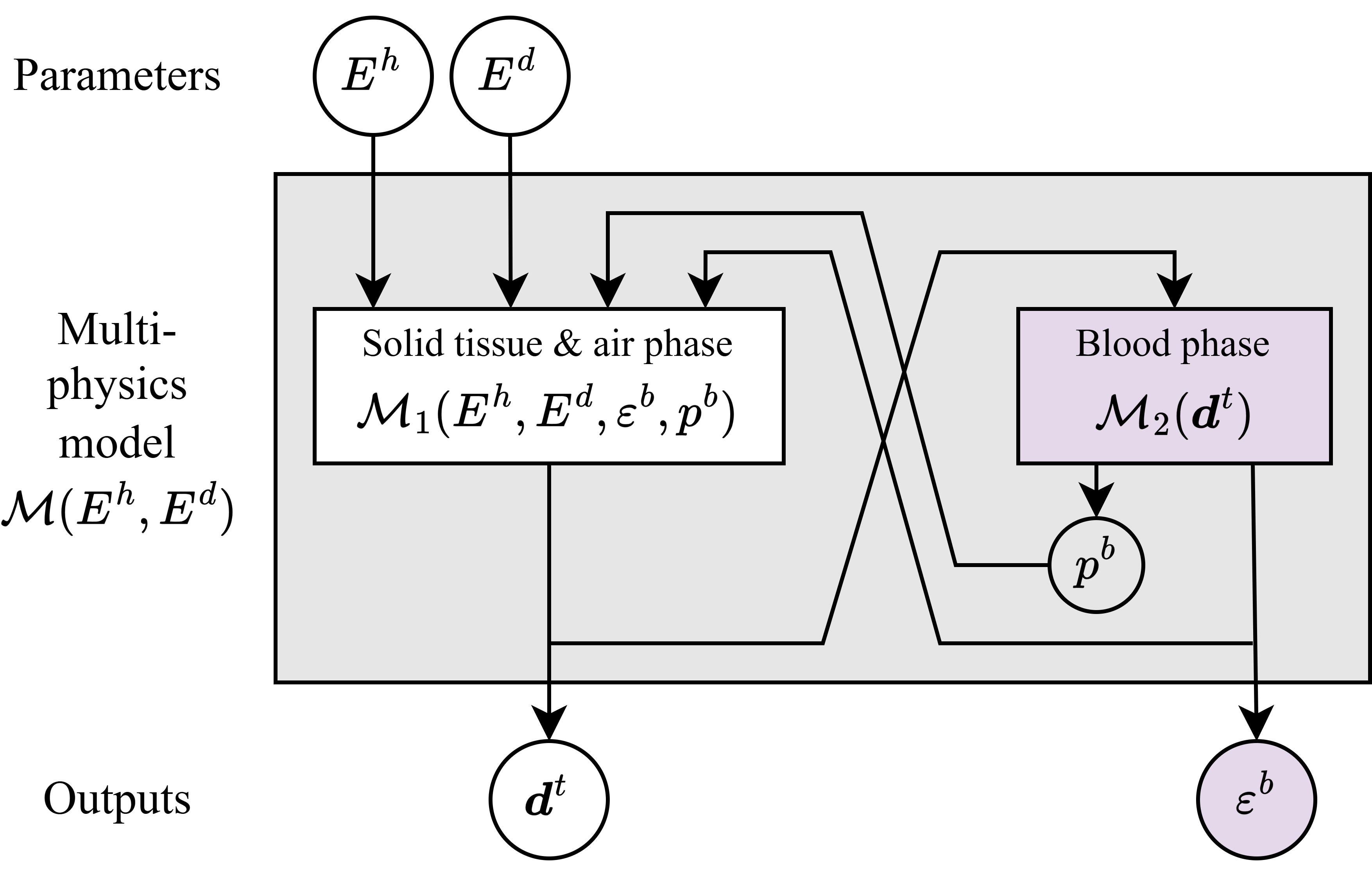}
    \caption{
        \emph{Left:} The domain of the mixed-dimensional multiphase porous medium model.
        The small rectangle on the right represents a diseased region of the porous lung tissue with a higher Young's modulus $E^d$.
        The remaining, healthy porous lung tissue has Young's modulus $E^h$.
        The porous domain is coupled to a discrete major airway and an artery on the top of the domain, and a vein at the bottom of the domain.
        \emph{Right:} The resulting fully coupled multiphase porous medium model $\M(E^h, E^d)$.
        Model $\M_1$ only considers the solid tissue with air in its pore space.
        We extend this model with blood as a second phase in the pore space.
        The resulting blood phase model $\M_2$ depends on $\M_1$ through the displacement of the tissue $\bm{d}^t$.
        The outputs of the blood phase model $\M_2$ (the blood volume fraction $\varepsilon^b$ and the blood pressure $p^b$) also impact the solid tissue and air phase model $\M_1$.
        Purple components are only added to the model to enhance the inverse analysis of $E^h$ and $E^d$, not the forward prediction of $\bm{d}^t$.}
    \label{fig:poro_domain_and_coupling}
\end{figure}
%%%%%%%%%%%%%%%%%%%%%%%%

\subsubsection{Forward model}

% Model introduction & 
% Purpose of the Demonstration / Why do we choose this model?
We model the lung tissue patch with the mixed-dimensional multiphase porous medium model presented in \cite{koeglmeier2025}.
The model describes the interaction of solid lung tissue with airways and blood vessels.
Larger airways and blood vessels are included in the model as discrete zero-dimensional networks.
Smaller respiratory and vascular units and the solid lung tissue are modelled by a homogenized porous medium.
In the porous medium, each of the resulting three physical fields (tissue $t$, air $a$, and blood $b$) is assigned a volume fraction $\varepsilon^t, \varepsilon^a$, and $\varepsilon^b$:
\begin{align}
    \underbrace{\varepsilon^t}_\text{tissue} + \underbrace{\varepsilon^a}_\text{air} + \underbrace{\varepsilon^b}_\text{blood} = 1 \eeq
    \label{eq:poro_volume_fractions}
\end{align}
We solve this multi-physics model for the unknown primary variables: the tissue displacement $\bd^t$, the air pressure $p^a$, and the blood pressure $p^b$.
The model is implemented and solved using our open-source multi-physics research code 4C \cite{fourc2025}.
The governing equations of the model are not essential for this work but are derived in detail in \cite{koeglmeier2025}.
We discretize these equations in space and time, resulting in a coupled system of nonlinear algebraic equations $\f^t$,  $\f^a$, and $\f^b$ for each of the three fields (compare Equation \ref{eq:nonlinear-system}):
\begin{align}
\begin{aligned}
    \f^t(\bd^t, \bp^a, \bp^b) = \bm{0} \ceq \\
    \f^a(\bd^t, \bp^a, \bp^b) = \bm{0} \ceq \\
    \f^b(\bd^t, \bp^a, \bp^b) = \bm{0} \eeq
    \label{eq:poro_model}
\end{aligned}
\end{align}
% Domain/Similarities & Differences to Köglmeier et al.
For the inverse problem at hand, we mostly adopt the geometry and the initial and boundary conditions of the quasi two-dimensional example from Section 3.1 in \cite{koeglmeier2025}.
Here, we only provide a brief summary of the model setup and describe the differences to \cite{koeglmeier2025}.
Our multiphase porous medium has a rectangular domain with dimensions $\SI{10}{mm} \times \SI{10}{mm}$ as shown on the left in Figure \ref{fig:poro_domain_and_coupling}.
The domain is divided into a healthy and a diseased region with differing Young's moduli $E^h$ and $E^d$.
We discretize the porous domain spatially with the finite element method using a regular mesh of $50 \times 50$ two-dimensional bilinear elements.
In contrast to \cite{koeglmeier2025}, we disregard the scalar transport of oxygen and carbon dioxide and the movement of the major airway, artery, and vein.
We further prescribe slightly higher pressures at the inlet of the major artery ($\smash{p^{\hat{b}}} = \SI{700}{Pa}$) and at the outlet of the major vein ($\smash{p^{\hat{b}}} = \SI{600}{Pa}$) to prevent capillary collapse at maximum air pressure.
In Section \ref{sec:model2_posteriors}, we compare the inverse results obtained using the three-field model described above with those from a simplified two-field model that excludes the blood field.
For this two-field model, we disregard the discrete major blood vessels (artery and vein) and assume that the smaller vascular units are included in the tissue volume fraction $\varepsilon^t$.

\paragraph{Coupling between the fields}
The three-field porous medium model with tissue, air, and blood is a fully coupled multi-physics model.
Figure \ref{fig:poro_domain_and_coupling} illustrates this coupling on the right.
The additional blood model $\M_2$ depends on the tissue and air fields through the tissue displacement $\bm{d}^t$
\footnote{For the model parameters selected here, the ratio between the air pressure and the blood pressure $p^a / p^b$ is always less than 1.
Thus, the evolution equation from \cite{koeglmeier2025} reduces to Equation \ref{eq:poro_evolution}.}:
\begin{align}
    \varepsilon^b =  
        \varepsilon^b_0 \cdot J^{k_J}
        \ \ \text{ with } J=\det\!\left(\bm{I} + \frac{\partial\bm{d}^t}{\partial\bm{X}}\right) \eeq
    \label{eq:poro_evolution}
\end{align}
This evolution equation takes into account the initial blood volume fraction at end-expiration $\varepsilon^b_0$, the determinant of the deformation gradient $J$ with $\bm{X}$ being the material coordinates, and the exponential parameter $k_J$.
This parameter $k_J$ determines the coupling strength from the displacement $\bd^t$ to the blood volume fraction $\varepsilon^b$.
Please refer to \cite{koeglmeier2025} for the coupling mechanism from the blood field back to the tissue and air fields.

\paragraph{Linearized system of equations}
We solve the fully coupled system of nonlinear equations from Equations \ref{eq:poro_model} using a fully monolithic solution algorithm with a single Newton-Raphson loop in each time step.
The linearized system of equations for the porous domain at an arbitrary time step and Newton iteration $n$ takes the form
\begin{align}
    \begin{bmatrix}
        \dfrac{\partial \f^t}{\partial \bd^t} & \dfrac{\partial \f^t}{\partial \bp^a} & \dfrac{\partial \f^t}{\partial \bp^b}\\
        \dfrac{\partial \f^a}{\partial \bd^t} & \dfrac{\partial \f^a}{\partial \bp^a} & \dfrac{\partial \f^a}{\partial \bp^b} \\
        \dfrac{\partial \f^b}{\partial \bd^t} & \dfrac{\partial \f^b}{\partial \bp^a} & \dfrac{\partial \f^b}{\partial \bp^b} 
    \end{bmatrix}
    ^{(n)}
    \begin{bmatrix}
        \Delta \bd^t \vphantom{\dfrac{\partial \f^t}{\partial \bp^a}}\\ 
        \Delta \bp^a \vphantom{\dfrac{\partial \f^a}{\partial \bp^a}}\\ 
        \Delta \bp^b \vphantom{\dfrac{\partial \f^b}{\partial \bp^a}}
    \end{bmatrix}
    ^{(n+1)}
    = -
    \begin{bmatrix}
        \f^t \vphantom{\dfrac{\partial \f^t}{\partial \bp^a}}\\ 
        \f^a \vphantom{\dfrac{\partial \f^a}{\partial \bp^a}}\\ 
        \f^b \vphantom{\dfrac{\partial \f^b}{\partial \bp^a}}
    \end{bmatrix}
    ^{(n)} \eeq
    \label{eq:linearized_poro_model}
\end{align}
None of the blocks in the system matrix are zero, which is characteristic of a fully coupled model.
See \cite{koeglmeier2025} for more details on the numerical implementation and solution process of the porous medium model.

% INVERSE METHODS

% Observations, noise, and prior
\subsubsection{Observations}
\label{sec:model2_observations}
\begin{figure}[]
    \includegraphics[width=\textwidth]{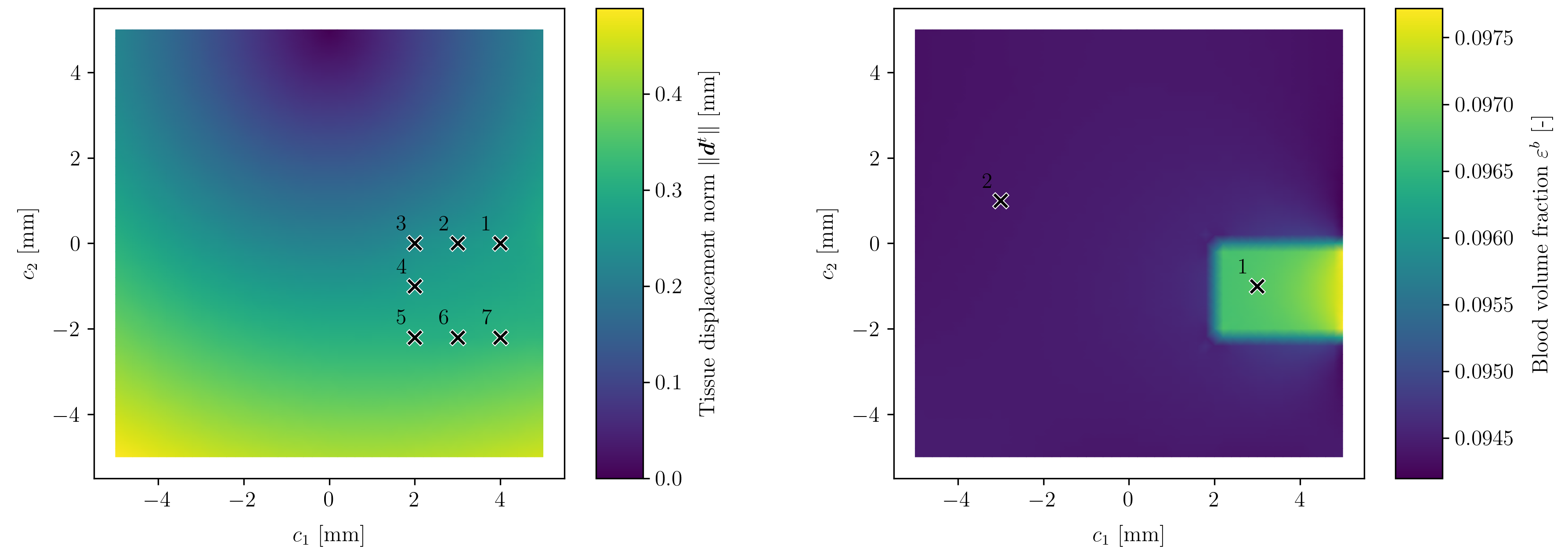}
    \caption{
        The solution fields of the tissue displacement $\bd^t$ (\emph{left}) and the blood volume fraction $\varepsilon^b$ (\emph{right}) for the ground-truth Young's moduli $E^h_\true$ and $E^d_\true$ at end-inspiration. 
        Observations are based on these solution fields and taken at the numbered locations marked with black crosses with white outlines.}
    \label{fig:obs_locations}
\end{figure}

We generated the synthetic observations based on an evaluation of the three-field porous medium model $\M(E^h, E^d)$ with the ground-truth parameters $E^h_\true = \SI{6}{kPa}$ and $E^d_\true = \SI{12}{kPa}$.
Starting the simulation from end-expiration, all observations were obtained after \SI{2.5}{s} at end-inspiration.
We extracted the displacements and blood volume fractions at the observation locations shown in Figure \ref{fig:obs_locations}.
For the displacements, we used $N_{\obs, 1} = 7$ observations and for the blood volume fractions, we used $N_{\obs, 2} = 2$ observations.
Note that each displacement observation is a vector with a component in $c_1$-direction and a component in $c_2$-direction.
To obtain the final observations, we added zero-mean Gaussian noise sampled from a Sobol' sequence \cite{sobol1967} according to Equation \ref{eq:noise}.
The variance of the noise $\sigma_j^2$ was derived from the signal-to-noise ratio $\SNR_j$ of each field $j$ (see Equation \ref{eq:snr}).

%%%%%%%%%%%%%%%%%%
%%%%%%%%%%%%%%%%%%

\subsubsection{Probabilistic model \& inference method}
For the prior distribution of the parameters of interest $E^h$ and $E^d$, we used a truncated normal distribution according to Table \ref{tab:params_model2}.
The observations were incorporated with the Gaussian multi-physics log-likelihood function presented in Equation \ref{eq:log-likelihood}\new{, assuming i.i.d.\! Gaussian noise with known variances}.
The likelihood variance $\sigma_j^2$ of each field $j$ was always identical to the variance used to sample the observational noise as described in Section \ref{sec:model2_observations}.
With this prior and likelihood, we evaluated the posterior on a $50\times50$ grid of $E^h$ and $E^d$.
The grid points in each dimension were evenly spaced with respect to the cumulative distribution function of the prior.
The model evaluations for the grid-based likelihood evaluations were scheduled in parallel using our open-source solver-independent multi-query framework QUEENS \cite{queens2025}.

% % % % % % % % % %
\begin{table}[b]
	\caption{Parameters of the prior used for the Bayesian inverse analyses using the porous medium model.}
	\label{tab:params_model2}
	\centering
	\begin{tabular}{lll}
        \addlinespace
		\toprule
		Parameter &
        Notation & 
        Value or Distribution \\
		\midrule
        Prior distribution & $p(\bx = [ E^h, E^d]^T )$ & $\mathcal{TN} \left( \bmu,\Sigma,\ba,\bb \right) $ \\
        Prior mean & $\bmu$ & $\left[\SI{7}{kPa}, \SI{15}{kPa}\right]^T$ \\
        Prior covariance & $\Sigma$ & $\diag\left((\SI{3}{kPa})^2, (\SI{10}{kPa})^2\right)$ \\
        Lower truncation bound & $\ba$ & $\left[\SI{0}{kPa}, \SI{0}{kPa}\right]^T$ \\
        Upper truncation bound & $\bb$ & $\left[\infty, \infty\right]^T$ \\
		\bottomrule
	\end{tabular}
\end{table}
% % % % % % % % % %

\subsubsection{Comparison of two single-physics and a multi-physics-enhanced Bayesian inverse analyses}
\label{sec:model2_posteriors}

\paragraph{Method}
% Method 1st figure
First, we compare the observations and results of two single-physics Bayesian inverse analyses (BIAs) and a multi-physics-enhanced BIA
\footnote{
    We use the term "single-physics BIA" to refer to a BIA that only considers observations of one physical field.
    This term is used regardless of whether the underlying forward model is a single-physics or a multi-physics model.
}.
The single-physics BIAs only consider observations of the tissue displacement $\bd^t=[d^t_1, d^t_2]$ with $\SNR_1 = 50$.
The first single-physics BIA uses the two-field model $\M_1$ (compare right plot in Figure \ref{fig:poro_domain_and_coupling}) that models air as the only phase present in the pore space of the solid tissue.
This model disregards the blood phase as a separate field.
The second single-physics BIA and the multi-physics-enhanced BIA use the fully coupled three-field model $\M$ that models both air and blood in the pore space of the solid tissue.
For both of these BIAs, we set the coupling parameter to $k_J =-\frac{2}{3}$.
In addition to the displacement observations, the multi-physics-enhanced BIA also includes observations of the blood volume fraction $\varepsilon^b$ with $\SNR_2=5.0\times10^4$.
To compute the relative increase in information gain (RIIG), we use the information gain of the single-physics BIA with the three-field model as the reference.

\paragraph{Results \& discussion}
% Results and Discussion 1st figure
Figure \ref{fig:model2_posteriors} shows the observations, posteriors, and information gains of the two single-physics BIAs versus the multi-physics-enhanced BIA.
The single-physics posterior based on the two-field model is slightly shifted towards lower values of $E^h$ compared to the single-physics posterior based on the three-field model.
We attribute this shift to the model error introduced by disregarding the blood phase and the resulting higher tissue volume fraction.
Regardless of the shift, both single-physics BIAs result in uninformative, widespread posteriors due to the low SNR of the displacement observations.
The information gains of both single-physics posteriors are comparable.
These similar information gains show that improving the forward model predictions by adding another physical field to the model has little effect on reducing uncertainty without incorporating additional observations from this field.
However, performing multi-physics-enhanced BIA and adding observations of the blood volume fraction $\varepsilon^b$ results in a considerably more concentrated posterior.
In particular, the uncertainty in the Young's modulus of the diseased region $E^d$ decreases.
This decrease in uncertainty is also reflected in the multi-physics information gain, which is considerably higher than the single-physics information gains.
The RIIG from the blood volume fraction observations is $\RIIG{\by_{1, \obs}, \by_{2, \obs}}=0.68$.
Even though we only used two additional blood volume fraction observations, the RIIG is still notable due to the high SNR of the blood volume fraction observations.

\subsubsection{Relative increase in information gain for varying coupling strength and signal-to-noise ratios}

%%%%%%%%%%%%%%%%%%
\begin{figure}
    \centering
    \includegraphics[width=\textwidth]{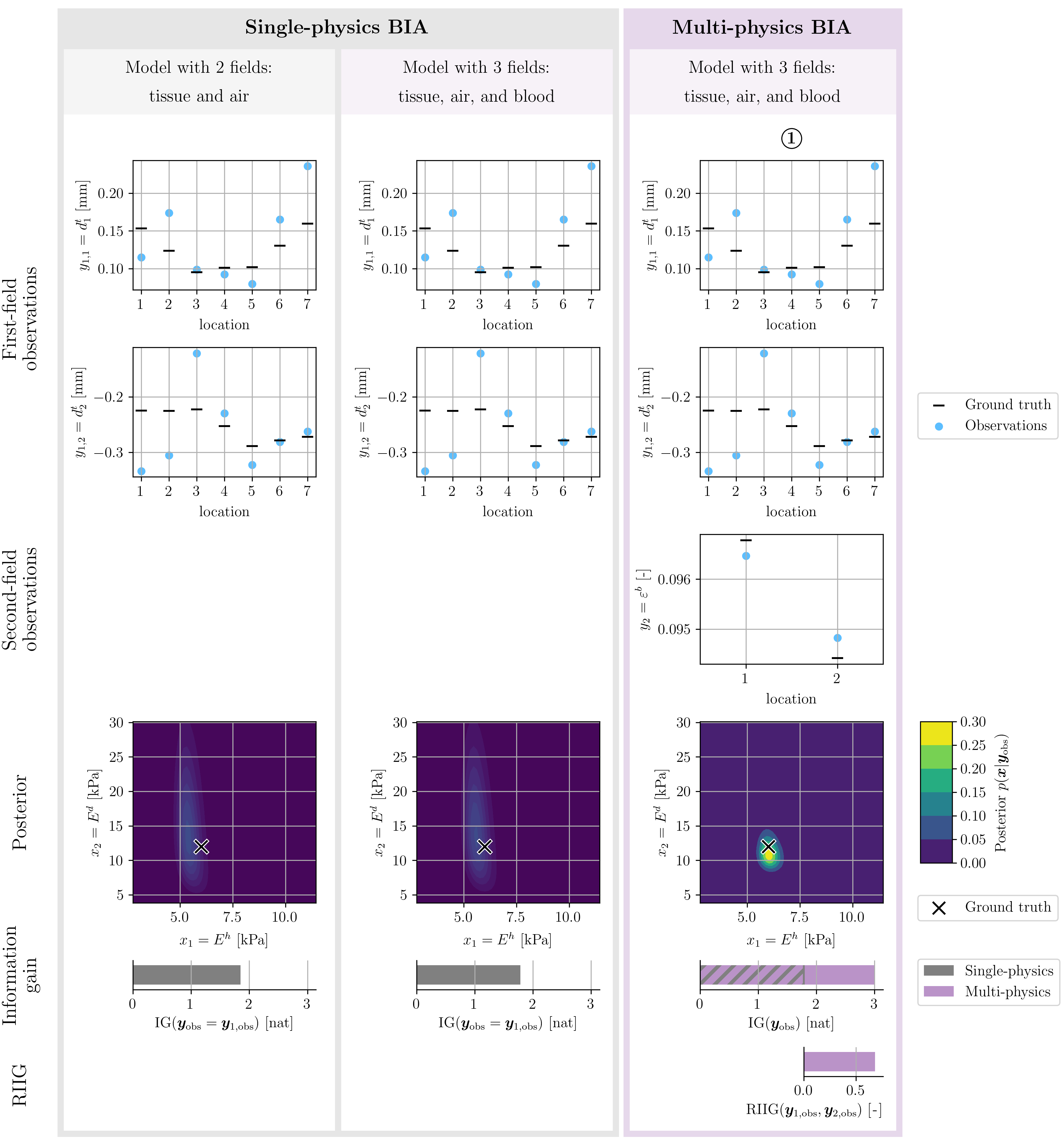}
    \caption{Observations and results of three Bayesian inverse analyses (BIAs) with different porous media models and observations.
    The two single-physics posteriors differ by a small shift in $E^h$ for the two-field model, but have a comparable large spread.
    Consequently, both single-physics posteriors have similar information gains.
    The multi-physics posterior is more concentrated and shows reduced uncertainty in $E^d$, although we only used two additional blood volume fraction observations.
    As a result, the multi-physics information gain is substantially higher than the single-physics information gains.
    The relative increase in information gain is $\RIIG{\by_{1, \obs}, \by_{2, \obs}} = 0.68$.}
    \label{fig:model2_posteriors}
\end{figure}
%%%%%%%%%%%%%%%%%%

%%%%%%%%%%%%%%%%%%
\begin{figure}
    \centering
    \vspace{-5pt}
    \begin{tikzpicture}
        \node[anchor=south west,inner sep=0] (img) at (0,0) 
            {\includegraphics[width=\textwidth]{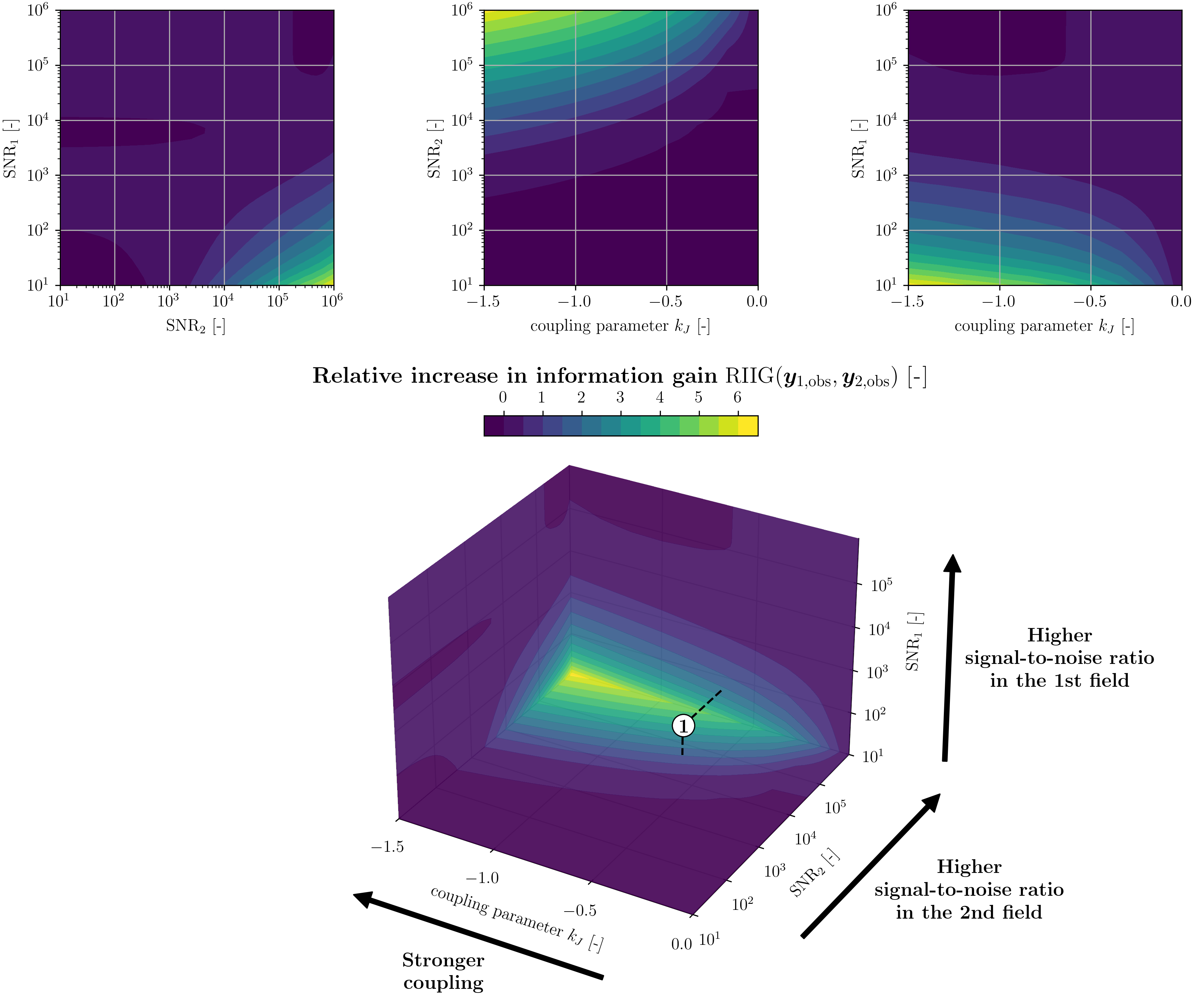}};

        \draw[arrows = {-Stealth[inset=0pt, length=6pt, angle'=60]}, line width=0.4mm, gray] (3.0,8.7) -- (5.2,6.7); % (x_end,y_end) -- (x_tip,y_tip)
        \draw[arrows = {-Stealth[inset=0pt, length=6pt, angle'=60]}, line width=0.4mm, gray] (14.15,8.7) -- (12.3,7.2);
        \node[anchor=south west,inner sep=0] (img) at (0,-4.5) 
            {\includegraphics[width=\textwidth]{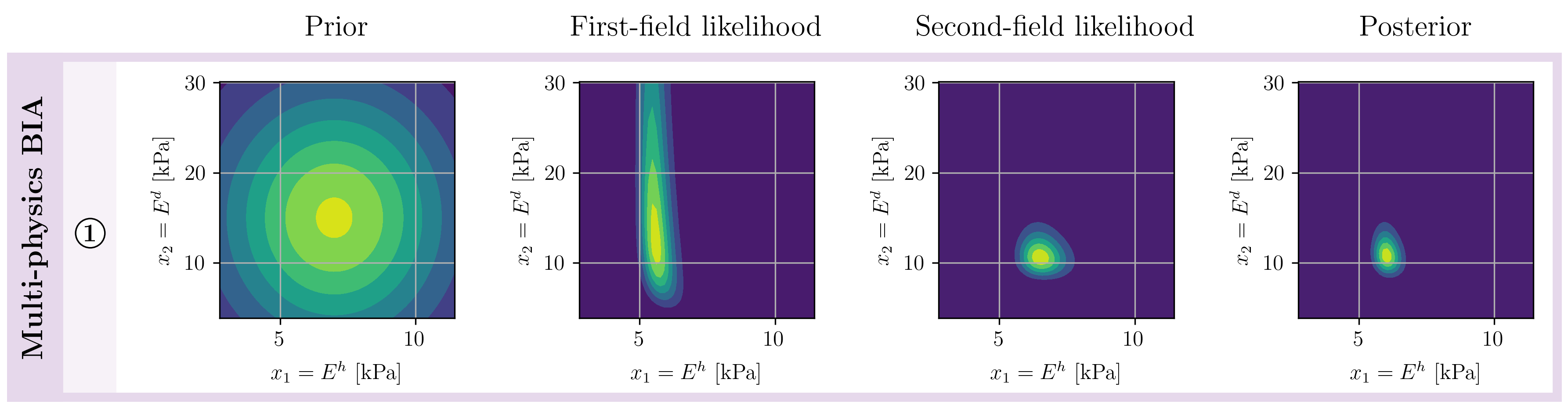}};
    \end{tikzpicture}
    \vspace{-10pt}
    \caption{
        The relative increase in information gain (RIIG) for $k_J=-1.5$ (\emph{top left}), $\SNR_1=10$ (\emph{top center}), and $\SNR_2=10^6$ (\emph{top right}) while varying the two remaining parameters.
        \new{\emph{Center:}} We also combine these three plots in a single, three-dimensional plot\delete{ (\emph{bottom})}. 
        \delete{The results demonstrate that for}\new{For} the example at hand, the RIIG from the single-physics to the multi-physics posterior is particularly high when the first field has a low SNR, the second field has a high SNR, and the coupling between the fields is strong\,\textsuperscript{6}.
        Even for weak coupling ($k_J \in [-0.5, -0.1]$), the RIIG is still notable if the $\SNR_1$ is sufficiently low and the $\SNR_2$ sufficiently high. 
        \new{\emph{Bottom:} The prior, likelihoods of each observed field $\by_1 = \bd^t$ and $y_2 = \varepsilon^b$, and posterior of the multi-physics-enhanced Bayesian inverse analysis (BIA) indicated in the center plot and already presented in Figure \ref{fig:model2_posteriors}.
        The likelihood of the second field considerably constrains the posterior.}
        \\
        \noindent\rule{0.311\textwidth}{0.4pt}\\
        \footnotesize$\hphantom{000}$\textsuperscript{6}
        Although mathematically implausible, the RIIG assumes slightly negative values for some combinations of $k_J$, $\SNR_1$, and $\SNR_2$ in Figure \ref{fig:model2_riig}.
        These negative values can be attributed to numerical errors in the information gain computation when a highly concentrated posterior is evaluated on a finite grid.
    }
    \label{fig:model2_riig}
\end{figure}
%%%%%%%%%%%%%%%%%%

\paragraph{Method}
% Method 2nd figure
In addition to the single multi-physics posterior and its \delete{information gain}\new{RIIG} in Figure \ref{fig:model2_posteriors}, we also analyze the RIIG for a range of first-field and second-field observations as well as for different coupling parameters $k_J$.
While we varied the number and SNR of the second-field observations in section \ref{sec:simple_model}, we now vary the SNRs of both the first-field and second-field observations and the coupling parameter $k_J$.
We evaluated the resulting RIIG on a $50\times50\times10$ grid of $\SNR_1$, $\SNR_2$, and $k_J$.
The grid points of the SNRs were log-spaced, and the grid points of $k_J$ were evenly spaced.

\paragraph{Results \& discussion}
% Results & Discussion 2nd figure
\new{The top and center plots in }Figure \ref{fig:model2_riig} show\delete{s} the resulting RIIG when varying these three parameters.
\new{Each point in these plots corresponds to a multi-physics-enhanced BIA.
This is also illustrated by the bottom plots in Figure \ref{fig:model2_riig}, which show the prior, the likelihoods of each observed field, and the multi-physics posterior at the indicated location in the center plot.
This posterior was already presented in Figure \ref{fig:model2_posteriors}.
The likelihood of the additionally observed second field intersects with the likelihood of the initially observed first field only in a small region of the parameter space.
This results in a concentrated posterior with corresponding RIIG of 0.68.}
The results \new{presented in the top and center plots} demonstrate that the RIIG is higher for a lower SNR in the first field and a higher SNR in the second field.
This is to be expected since a lower SNR in the first field means that observations of this field are less informative, resulting in an increasingly uninformative single-physics posterior.
As a result, the single-physics information gain decreases, which increases the RIIG.
Conversely, a higher SNR in the second field means that observations of this field are more informative.
This increases the multi-physics information gain, which in turn increases the RIIG.
Figure \ref{fig:model2_riig} further illustrates that the RIIG is higher the stronger the coupling between the two fields.
This observation also aligns with expectations since a stronger coupling implies that the second field is influenced more strongly by the first field, which depends on the parameters of interest.
We also observe that the RIIG is zero when the coupling parameter $k_J$ is zero.
A coupling parameter of zero means that the second field does not depend on the first field.
Hence, the second field contains no information about the parameters of interest, and the RIIG is zero.
Furthermore, the RIIG is still notable even for weak coupling between the fields ($k_J \in [-0.5, -0.1]$) if the $\SNR_1$ is sufficiently low and the $\SNR_2$ sufficiently high.
In cases where the $\SNR_2$ is not high enough, we assume that a sufficiently large number of second-field observations can compensate for this lack of information and increase the RIIG as shown in section \ref{sec:simple_model}.
Our findings indicate that although strong coupling yields a higher RIIG, observations of a weakly coupled second physical field can still enhance a Bayesian inverse analysis.

%%%%%%%%%%%%%%%%%%%%%%%%%%%%%%%%%%%%%%%%%%%%%%%%%%%%%%%%%%%%%%%%%%%%%%%%%%%%%%%%%%%%%%%%%%%%%%%%%%%
\FloatBarrier
\section{Conclusion}
\label{sec:conclusion}

This work proposes \delete{the approach of}\new{a novel approach that we denote as} multi-physics-enhanced Bayesian inverse analysis.
\new{This approach leverages observations of additional physical fields to identify parameters of single-physics computational models, avoiding the costly redesign of experimental setups to enhance single-physics observations.}
We demonstrate that \delete{this}\new{the multi-physics-enhanced} approach can substantially reduce the uncertainty in \delete{a previously}\new{an} uninformative posterior distribution\delete{ by adding observations of additional physical fields}.
\delete{This}\new{We quantify this} reduction in uncertainty \delete{is quantified }by evaluating the information gain from the prior to the posterior distribution.
To assess the effect of the multi-physics observations, \delete{we compared}\new{our work is the first to compare} the information gain with and without observations from a second physical field.
Our results show that even a few or noisy additional observations of a second physical field can reduce uncertainty considerably.
Even additional observations from just a one-way coupled or weakly coupled physical field resulted in a notable uncertainty reduction.
\new{The uncertainty reduction through the second-field observations was generally higher the lower the signal-to-noise ratio (SNR) in the first field, the higher the SNR and the number of observations in the second field, and the stronger the coupling between the two fields.}
Overall, our findings highlight the largely neglected potential of multi-physics observations to enhance \new{(}Bayesian\new{)} inverse analysis.
This considerable potential can be \delete{leveraged}\new{exploited} with minimal effort in the frequent case where multi-physics observations are easily obtainable and a corresponding multi-physics model is available or easy to implement.
This implementation effort is particularly low \new{when the multi-physics model used for the inverse analysis has either one-way coupled or uncoupled physical fields}\delete{for one-way coupled or uncoupled multi-physics models}.
Building on our findings, we encourage experimenters, modelers in academia and industry, and developers of inverse methods to leverage the potential of additional physical fields and address previously intractable inverse problems.

%%%%%%%%%%%%%%%%%%%%%%%%%%%%%%%%%%%%%%%%%%%%%%%%%%%%%%%%%%%%%%%%%%%%%%%%%%%%%%%%%%%%%%%%%%%%%%%%%%%
\section*{Acknowledgements}

We gratefully acknowledge financial support by BREATHE, an ERC-2020-ADG project, grant agreement ID 101021526.

\section*{Declaration of generative AI and AI-assisted technologies in the manuscript preparation process}

During the preparation of this work, the authors used ChatGPT (GPT-3.5, GPT-4o, GPT-5) and GitHub Copilot (GPT-4.1) in order to improve language and readability. After using these tools, the authors reviewed and edited the content as needed and take full responsibility for the content of the published article.

\printbibliography

\appendix

\FloatBarrier
%%%%%%%%%%%%%%%%%%%%%%%%%%%%%%%%%%%%%%%%%%%%%%%%%%%%%%%%%%%%%%%%%%%%%%%%%%%
\section{Derivation of the one-way coupled electromechanical model}
\label{ap:simple_model}

\subsection*{Mechanical model}
\renewcommand{\arraystretch}{1.2}

The mechanical model in Section \ref{sec:simple_model} considers a cube made of the material of interest.
The material is isotropic and homogeneous and is modelled by a Saint Venant-Kirchhoff material model.
The cube has side lengths $l_0 = \SI{10}{mm}$ and thus a square cross-sectional area $A_0=\left(l_0\right)^2$ in its undeformed state.
Starting from this state, a force $F$ acts on the upper surface of the cube in $c_1$-direction while the cube's lower surface is fixed in $c_1$-direction.
As a result, the cube elongates to length $l=l_0 + d$ with $d$ being the displacement of the cube's upper surface.
The lateral surfaces of the cube normal to the $c_2$-direction are free to contract, while the lateral surfaces normal to the $c_3$-direction are fixed in $c_3$-direction.
Figure \ref{fig:simple_model} displays this setup.

The deformation gradient $\bm{F}$ of the deformed cube is given by
\begin{align}
    \bm{F} = \begin{bmatrix}
        F_{11} =  \frac{l}{l_0} & 0 & 0 \\ 
        0 & F_{22} & 0 \\ 
        0 & 0 & 1
    \end{bmatrix} \eeq
\end{align}
The Green-Lagrange strain tensor $\bm{E}$ for this deformation is given by
\begin{align}
    \bm{E} =
    \begin{bmatrix}
        E_{11} = \frac{1}{2} (\frac{l^2}{l_0^2} - 1) & 0 & 0 \\ 
        0 & E_{22} = \frac{1}{2} (F_{22}^2 - 1) & 0 \\ 
        0 & 0 & 0
    \end{bmatrix} \eeq
    \label{eq:gl_strain}
\end{align}
Since we assume a Saint Venant-Kirchhoff material model, the second Piola-Kirchhoff stress tensor $\bm{S}$ is given by
\begin{align}
    \bm{S} =
    \begin{bmatrix}
        S_{11} & 0 & 0 \\ 
        0 & S_{22} = 0 & 0 \\ 
        0 & 0 & S_{33}
    \end{bmatrix} 
    := \lambda \text{tr}(\bm{E}) \bm{I} + 2 \mu \bm{E}  \ceq
\end{align} 
where $\lambda$ and $\mu$ are the Lamé parameters. We obtain $S_{22} = 0$ since the cube is free to move in $c_2$-direction and does not experience any surface traction in this direction.
From this constraint, we derive the relation between $E_{11}$ and $E_{22}$:
\begin{align}
\begin{aligned}
    S_{22} & = \lambda \text{tr}(\bm{E}) + 2 \mu E_{22}\\
    & = \frac{\nu E}{(1 + \nu)(1-2\nu)}\left(E_{11} + E_{22}\right) + \frac{E}{1 + \nu} E_{22} \\
    & = \frac{\nu E}{(1 + \nu)(1-2\nu)} E_{11} + \frac{(1 - \nu) E}{(1 + \nu)(1-2\nu)} E_{22}\\
    & := 0
\end{aligned}
\end{align}
and thus
\begin{align}
    E_{22} = - \frac{\nu}{1-\nu} E_{11} \eeq
    \label{eq:strain_relation}
\end{align}
We then use this relation to compute $S_{11}$:
\begin{align}
\begin{aligned}
    S_{11} &= \lambda \text{tr}(\bm{E}) + 2 \mu E_{11} \\
    & = \frac{\nu E}{(1 + \nu)(1-2\nu)}E_{11} \left(1 - \frac{\nu}{1-\nu}\right) + \frac{E}{1 + \nu} E_{11} \\
    & = \left( \frac{\nu E}{(1 + \nu)(1-\nu)} + \frac{E}{1 + \nu}\right) E_{11} \\
    & = \frac{E}{1 - \nu^2} E_{11} \\
    & = \frac{E}{2(1 - \nu^2)} \left(\frac{(l_0 + d)^2}{l_0^2} - 1\right) \\
    & = \frac{E (2 l_0 d + d^2)}{2 l_0^2 (1-\nu^2)} \eeq
\end{aligned}
\end{align}
We assume the external force $F$ distributes evenly across the cube's deformed upper surface $A$.
Hence, the external traction $\bm{t}$ in the current configuration on this surface becomes
\begin{align}
    \bm{t} = \begin{bmatrix}\frac{F}{A} \\ 0 \\ 0 \end{bmatrix} \eeq
\end{align}
The external traction $\bm{T}$ in the initial configuration for a constant deformation gradient is then given by
\begin{align}
    \bm{T} = \bm{F}^{-1} \cdot \frac{A}{A_0} \bm{t} =
    \begin{bmatrix}
        \frac{l_0}{l} & 0 & 0 \\ 
        0 & F_{22}^{-1} & 0 \\ 
        0 & 0 & 1
    \end{bmatrix} \cdot \begin{bmatrix}\frac{F}{A_0} \\ 0 \\ 0 \end{bmatrix}
    = \begin{bmatrix}\frac{F}{l_0(l_0 +d)} \\ 0 \\ 0 \end{bmatrix} \eeq
\end{align}
Assuming a quasi-static deformation and no gravitational effects, we then formulate the balance of forces on the cube's upper surface with the outward normal vector $\bm{N}$:
\begin{align}
    \bm{S} \cdot \bm{N} - \bm{T}
    = \begin{bmatrix}
        S_{11} & 0 & 0 \\ 
        0 & 0 & 0 \\ 
        0 & 0 & S_{33}
    \end{bmatrix} \cdot 
    \begin{bmatrix}
         1 \\ 0 \\ 0
    \end{bmatrix}
    -
    \begin{bmatrix}
         \frac{F}{l_0(l_0 +d)} \\ 0 \\ 0
    \end{bmatrix}
    =
    \begin{bmatrix}
        S_{11} - \frac{F}{l_0(l_0 +d)} \\ 0 \\ 0
    \end{bmatrix}
    := \bm{0} \eeq
\end{align}
Only looking at the first component of this equation, we obtain the following relation:
\begin{align}
    \frac{E (2 l_0 d + d^2)}{2 l_0^2 (1-\nu^2)} - \frac{F}{l_0(l_0 +d)} = 0 \eeq
\end{align}
After rearranging this equation, we derive the following non-linear relation between the applied force $F$ and the displacement $d$:
\begin{align}
    f_1(d) = 2 l_0^2 d + 3 l_0 d^2 + d^3 - \frac{2 F l_0}{E} \left(1 - \nu ^2\right) = 0 \eeq
\end{align}

\subsection*{Electrical model}
The electric current $I$ through the cube is equal to the ratio of the potential difference $U$ between the cube's opposing surfaces and the cube's electrical resistance $R$. This resistance $R$ depends on the electrical resistivity $\rho$ of the cube's material, the current length $l$ of the cube, and its current cross-sectional area $A$:
\begin{align}
   \frac{U}{I} = R  = \frac{\rho l}{A} \eeq
   \label{eq:resistance}
\end{align}
We derive the current cross-sectional area $A$ from the determinant of the deformation gradient $\bm{F}$, which quantifies the change in volume of the cube due to the deformation:
\begin{align}
\begin{aligned}
    V &= \det(\bm{F}) V_0 \\
    A l &= F_{11} F_{22} l_0^3
\end{aligned}
\end{align}
and thus
\begin{align}
    A = \frac{ \frac{l}{l_0} F_{22} l_0^3}{l} = l_0^2 F_{22} \eeq
\end{align}
The deformation gradient entry $F_{22}$ is computed by solving the Green-Lagrange strain entry $E_{22}$ from Equation \ref{eq:gl_strain} for $F_{22}$ and plugging in the relation between $E_{22}$ and $E_{11}$ from Equation \ref{eq:strain_relation}:
\begin{align}
\begin{aligned}
    F_{22} &= \sqrt{2 E_{22} + 1}  \\
    &= \sqrt{- 2 \frac{\nu}{1-\nu} E_{11} + 1} \\
    &= \sqrt{- \frac{\nu}{1-\nu} (\frac{(l_0 + d)^2}{l_0^2} - 1) + 1}  \\
    &= \sqrt{- \frac{\nu}{1-\nu} \frac{2 l_0 d + d^2}{l_0^2} + 1} \\
    &= \frac{1}{l_0}\sqrt{ l_0^2 - \frac{\nu}{1-\nu} (2 l_0 d + d^2)}  \eeq
\end{aligned}
\end{align}

Rearranging Equation \ref{eq:resistance} and subsequently plugging in the derived expressions for $A$ and $F_{22}$ yields the non-linear relation between the displacement $d$ and the electric current $I$:
\begin{align}
\begin{aligned}
    f_2(d, I) &= \rho l I - U A \\
    &= \rho (l_0 + d) I - U l_0^2 F_{22} \\
    & = \rho (l_0 + d) I - U l_0 \sqrt{ l_0^2 - \frac{\nu}{1-\nu} (2 l_0 d + d^2)} \\
    & = 0 \eeq
\end{aligned}
\end{align}

\end{document}